\documentclass[aps,amsmath,superscriptaddress,amssymb,twocolumn]{revtex4-2}  
\usepackage{color}
\usepackage{hyperref}
\usepackage{enumitem}
\usepackage{graphicx,subfigure}
\usepackage[normalem]{ulem}
\usepackage{lineno}
\usepackage{booktabs}
\usepackage{balance}
   \usepackage{array}
    \usepackage{multirow}
    \usepackage{hhline}
    \usepackage{makecell}
     \usepackage[symbol]{footmisc}
\newif\ifHighlitedChanges
\def\ifHighlitedChanges{\iftrue}
\ifHighlitedChanges
   
  \def\STRIKE#1{{\color{red}\sout{#1}}}
\else
  
  \def\STRIKE#1{\relax}
\fi

\newif\ifHighlitedChanges
\def\ifHighlitedChanges{\iftrue}
\ifHighlitedChanges

\else

\fi

\begin{document}
\title{Microscopic pathways of transition from low-density to high-density amorphous phase of water}
\author{Gadha Ramesh}
\affiliation{Department of Physics, Indian Institute of Science Education and Research (IISER) Tirupati, Tirupati, Andhra Pradesh 517507, India}
\author{Ved Mahajan}
\thanks{Current Address: Technical University of Darmstadt, Darmstadt 64289 , Germany}
\affiliation{Department of Physics, Indian Institute of Science Education and Research (IISER) Tirupati, Tirupati, Andhra Pradesh 517507, India}
\author{Debasish Koner}
\affiliation{Department of Chemistry, Indian Institute of Technology Hyderabad, Hyderabad, Telangana 502285, India}
\author{Rakesh S. Singh}
\email{rssingh@iisertirupati.ac.in}
\affiliation{Department of Chemistry, Indian Institute of Science Education and Research (IISER) Tirupati, Tirupati, Andhra Pradesh 517507, India}
%\date{\today}
%%%%%%%%%%%%%%%%%%%%%%%%%%%%%%%%%%%%%%%%%%%%%%%%%%%%%
\begin{abstract}
Much attention has been devoted to understanding the microscopic pathways of phase transition between two equilibrium condensed phases (such as, liquids and solids). However, the microscopic pathways between non-equilibrium, non-diffusive amorphous (glassy) phases still remain poorly understood. In this work, we have employed computer simulations, persistence homology (a tool rooted in topological data analysis), and machine learning to probe the microscopic pathway of pressure-induced non-equilibrium transition between the low- and high-density amorphous (LDA and HDA, respectively) ice phases of TIP4P/2005 and ST2 water models. Using persistence homology and machine learning, we introduced a new order parameter that unambiguously identifies the LDA and HDA-like local environments. The system transitions continuously and collectively in the order parameter space via a pre-ordered intermediate phase during the compression of the LDA phase. The local order parameter susceptibilities show a maximum near the transition pressure ($P^*$) --- suggesting maximum structural heterogeneities near $P^*$. The HDA-like clusters are structurally ramified and spatially delocalized inside the LDA phase near the transition pressure. We also found the manifestations of the first-order low-density to high-density liquid transition in the sharpness of the order parameter changes during the LDA to HDA transition. We have further investigated the (geometrical) structures and topologies of the LDA and HDA ices formed via different protocols and also studied the dependence of the microscopic pathway of phase transition in the order parameter space on the protocol followed to prepare the initial LDA phase. Finally, the method adopted here to study the microscopic pathways of transition is not restricted to the system under consideration and provides a robust way of probing phase transition pathways involving any two condensed phases at both equilibrium and out-of-equilibrium conditions.  
\end{abstract}

\maketitle

\section{\label{sec:levelI}Introduction}
Water's anomalous properties make it fascinating, and its rich polymorphic behavior --- which is the ability to exist in multiple crystalline or glassy (amorphous) states --- adds even more intrigue to research on this subject~\cite{wateranomalydebenedetti2021, wateranomalybagchi2013, pablo_rev_2003}. Several experimental studies have shown that the thermodynamic response functions of water increase as the temperature decreases and appear to diverge as the temperature is further decreased below its melting temperature~\cite{wateranomalyangell1973, wateranomalyhare1987, wateranomalyspeedy1976, wateranomalytombari1999heat}. One of the theoretical scenarios proposed to interpret this anomalous divergence-like behavior posits the existence of a liquid-liquid critical point (LLCP)~\cite{LLCPpoole1992phase}. According to the LLCP scenario, water is believed to exist in two different forms, namely, low-density and high-density liquids (LDL and HDL, respectively), which are separated by a hypothetical first-order liquid-liquid transition (LLT) line ending at the LLCP~\cite{LLCPpoole1992phase}. It is extremely difficult, however, to unambiguously prove or disprove this hypothesis through experiments because ice nucleates rapidly before any measurement can be made on deeply supercooled water. Hence, over the years computer simulations have become an indispensable tool to study the thermodynamic behavior of water under extreme conditions and understand the precise origin of water's anomalies~\cite{LLCPpalmer2018advances, twostrucutresgallo2016water, LLCPdebenedetti2020second, twostrucutresshi2020direct, twostrucutresshi2020anomalies, liu2009low, bagchi_2011, twostrucutressingh2016two, twostrucutressingh2016two, singh_tip4p_2017, singh_tip4p_2017_2, singh_jcp_2022, pablo_prl_2022, pablo_jpcl_2022, molinero_jpcb_2022}. 

In the LLCP scenario, low-density amorphous (LDA) and high-density amorphous (HDA) phases are believed to be the glassy (kinetically-arrested) counterparts of the LDL and HDL phases~\cite{LDAHDAmishima1998}. Many experimental and simulation studies have detected a correspondence between the liquid and the glassy counterparts~\cite{connectionmcmillan1965vitreous, connectionjohari1987glass, connectionhandle2012relaxation, connectionamann2013water, connectionperakis2017diffusive, giovambattista2021liquid, karina2022infrared, foffi2021structure, fausto_2020}.  In addition, some experimental studies have suggested the nature of the compression-induced LDA to HDA transition to be first-order-like based on a seemingly discontinuous change in density and reversibility with hysteresis~\cite{mishima1985apparently, mishima_jcp_1994, nelmes_2006, yoshimura_2007}. Furthermore, recent computational studies have revealed remarkable similarities between the LDA and HDA ices and corresponding associated liquid phases through the potential energy landscape~\cite{giovambattista2016potential}, suggesting again a first-order-like LDA to HDA transition. Despite many experimental and computational studies indirectly suggesting the LDA to HDA transition a first order-like~\cite{giovambattista2012interplay}, due to the lack of any direct evidence, the precise nature of this transition still remains a matter of debate. This also brings in an ambiguity over the observations of whether LDA and HDA phases are structurally-arrested analog of the thermodynamically distinct LDL and HDL phases, or on a more general note, whether the nature of the LDA to HDA transition has any implications for our understanding of supercooled water's anomalous behavior. 

In thermal systems, order parameter hysteresis on reversibility is a consequence of the existence of metastability and suggests a first-order transition. However, the correspondence between the hysteresis and metastability (or, the nature of phase transition) might not be directly applicable to the transition between two non-equilibrium glassy phases. In the case of LDA and HDA phases, the hysteresis in the order parameter could be a consequence of the fact that the system is tracing different pathways in the configuration space during the compression and decompression. Also, for equilibrium systems, a first-order phase transition often involves discontinuous order parameter change. However, for a compression-induced transition between two non-diffusive (kinetically-arrested) phases, a discontinuous order parameter change might not always be possible due to its dependence on the compression rate. Hence, it would be challenging to establish the nature of the transition to be first-order-like or not solely based on the sharpness of the order parameter change, or, the behavior of (global) order parameter susceptibilities in the vicinity of the transition point. Therefore, the thermodynamic markers for the nature of the phase transition would not be directly relevant for these non-equilibrium transitions. Furthermore, in recent computational studies, it has been observed that the sharpness of the transition between LDA to HDA is correlated with the depth of the initial LDA sample in the PEL~\cite{PELgiovambattista2017influence}. Slower cooling and compression allow the system to explore deeper regions of the potential energy landscape. Hence, the protocol followed to prepare the initial LDA samples is also expected to influence the LDA to HDA transition pathways in the configuration space~\cite{wong2015pressure, handle2019glass, martelli2017large}. 

Considering the above limitations imposed by the non-ergodic nature of the system, it would be, therefore, desirable to understand the nature of phase transition by carefully probing its microscopic pathway --- whether it follows nucleation and growth-like mechanism or not --- and its dependence on the protocol followed to prepare the initial LDA phase. We would also like to emphasize here that probing the microscopic path of transition requires unambiguous identification of the growing new phase (say, HDA phase) and old parent phase (say, LDA phase) at molecular scales, which is an extremely difficult task as both these phases (LDA and HDA) lack global order (unlike liquid to solid transition). This could be the reason that surprisingly few attempts have been made to probe the microscopic pathway of LDA to HDA transition using computer simulations~\cite{mollica2022decompression, vmolineroFOPT, tanaka_natcom_2024}.

The short-range order (SRO) in amorphous ices, which describes the atomic configurations typically up to the first neighbor shell, has been extensively investigated to characterize the local structural features of the LDA and HDA phases~\cite{sro_exp_1, fausto_2020, foffi2021structure}. It has also been reported recently that the LDA and HDA phases exhibit an intriguing hyperuniform behavior~\cite{martelli2017large, formanek2023molecular} --- an anomalous suppression of large-scale density fluctuations compared to simple liquids~\cite{stillinger_hyperuniform}. It has become more evident now that larger-scale structures beyond neighboring atoms, referred to as medium-range order (MRO), play a more crucial role than SRO in glasses~\cite{sheng_mro_nat_2006, tanaka_prl_mro_2007, nakamura2015persistent, sastry_prl_mro_2018, tanaka_nat_rev_2019, hong2019medium}. That is, one needs to probe the many-body correlations involving particles beyond the first neighbor shell to understand the microscopic structure of glasses. To describe the many-body local structures in fluid and glassy systems, bond-orientational order parameters, and topological data analysis methods, such as persistence homology (PH), have recently been used~\cite{nakamura2015persistent, hong2019medium, hiraoka2016hierarchical}. In the PH analysis, an atomic configuration is considered a point cloud and is used to systematically extract the structural and topological features present in the system.

In this study, we performed molecular dynamic simulations (MD) with TIP4P/2005~\cite{abascal2005general} and ST2~\cite{stillinger1974improved} water to generate the LDA and HDA ices and probe the microscopic pathways of pressure-induced transition between them. The LDA and HDA structures, and phase behavior in general, for these two water models, are relatively well studied~\cite{wong2015pressure} compared to other water models. The ST2 water exhibits a readily accessible LLT line in computer simulations~\cite{LLCPpoole1992phase, pablo_nat_2014} compared to other rigid water models, and hence, it is possible to directly get the kinetically-arrested LDA and HDA phases by rapid cooling of the corresponding equilibrium LDL and HDL phases (unlike the TIP4P/2005 water where the LLT line is buried in the deep supercooled region of the phase plane and is not that easily accessible~\citep{LLCPdebenedetti2020second}). As discussed earlier, a structural order parameter capable of unambiguously identifying the LDA- and HDA-like local environments is a prerequisite for probing the microscopic mechanism of transition. Here, using PH and machine learning, we have introduced a new order parameter to unambiguously identify the LDA and HDA-like local environments. Using this order parameter, along with the locally-averaged bond-orientational and local density as alternate order parameters, we have carefully probed the microscopic LDA to HDA transition pathway. We have further explored the dependence of the (microscopic) transition pathways on the protocol followed to prepare the initial LDA phase.       

The structure of the rest of the paper is as follows. In Section~\ref{sec:level1} the computational protocols followed for preparing the glasses using the TIP4P/2005~\cite{abascal2005general} and ST2~\cite{stillinger1974improved} water models are described. In Section~\ref{subsec:21}, we discuss the global structural and topological features of the LDA and HDA ices prepared following different protocols. The (local) order parameters to identify the LDA and HDA-like local environments are discussed in Section~\ref{subsec:22}. The evolution of the system in the order parameter space during the compression of the LDA phase and decompression of the HDA phase is discussed in Section~\ref{subsec:23}. Section~\ref{subsec:24} probes the possibility of nucleation and growth-like mechanism of the transition between the amorphous phases by employing clustering analysis, and Section~\ref{subsec:25} contains our study on the dependence of the (microscopic) transition path on the protocol followed to prepare the initial LDA phase. The major conclusions from this work are summarized in Section~\ref{sec:c}. 
\begin{figure}
    \includegraphics[scale=0.35]{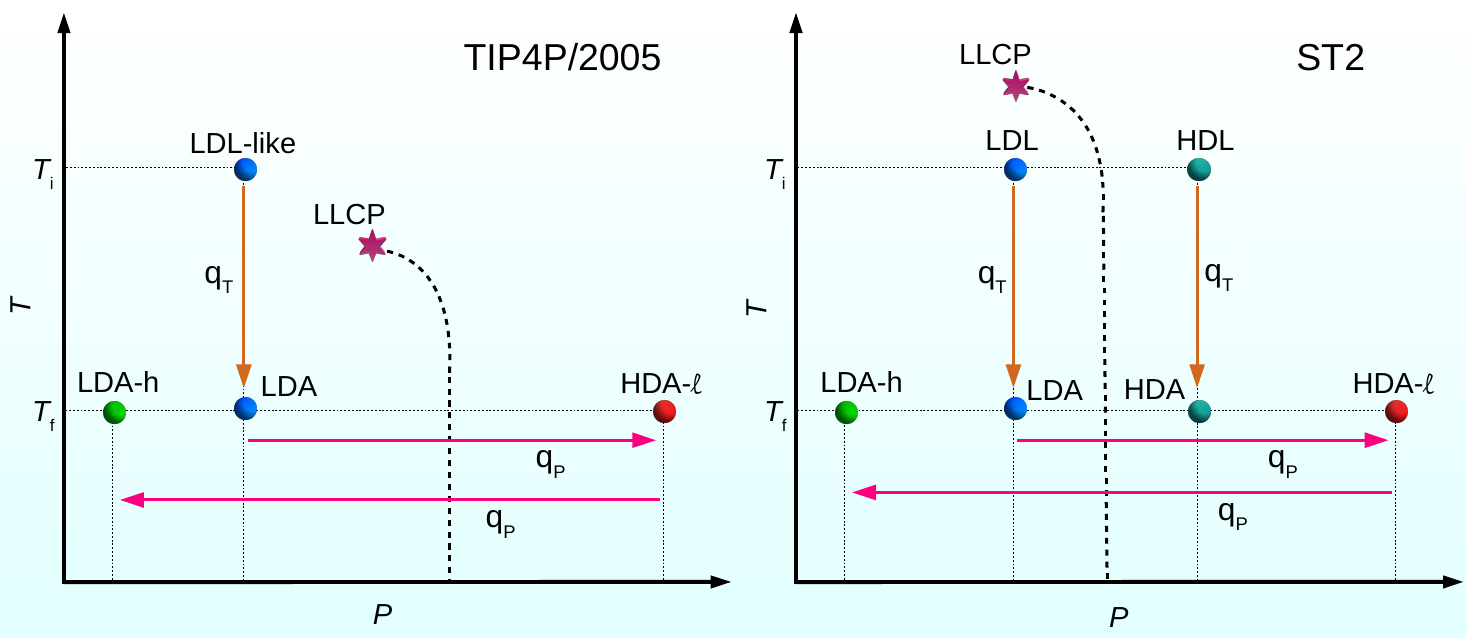}
    \caption{A schematic phase diagram showing the protocols followed to generate the LDA and HDA configurations for the TIP4P/2005 (left) and the ST2 (right) water. For TIP4P/2005 water, the equilibrated liquid water configurations at a supercritical temperature with respect to the liquid-liquid critical temperature ($T > T_c \sim 182$ K) were isobarically quenched to $80$ K to generate the LDA configurations. For the ST2 water, however, we have quenched the thermally equilibrated LDL phase at a subcritical temperature ($T < T_c \sim 247$ K) to get the corresponding LDA configurations. For both the water models, the LDA configurations were then isothermally compressed to get the corresponding HDA structures, named HDA-$l$. The HDA-$l$ configurations were isothermally decompressed to get the corresponding low-density configurations, named LDA-$h$ (solid pink lines). For the ST2 water, we also obtained the HDA phase from the thermally equilibrated HDL phase by quenching it isobarically.}
    \label{phase_diagram}
\end{figure}
\section{\label{sec:level1}Water Models and Simulation Details}
We have used two water models --- TIP4P/2005~\cite{abascal2005general} and ST2~\cite{stillinger1974improved}. For the TIP4P/2005 water, we performed MD simulations in the isothermal-isobaric ($NPT$) ensemble with GROMACS $5.1.2$~\cite{gromacs} on a system consisting of $N = 4000$ molecules in a cubic box. The short-range part of the interaction potential was truncated at $1.0$ nm, and long-range corrections were applied to the short-range interaction for energy and pressure. The Coulombic interaction was truncated at $1.0$ nm, and the particle mesh Ewald (PME)~\cite{darden1993particle} was used to compute the long-range contributions to the electrostatic interaction. Trajectories were propagated using a $2$ fs time step, and Nose-Hoover thermostat~\cite{nose, hoover} was used to maintain the constant temperature with a relaxation time of $0.4$ ps. The constant pressure was maintained using Parrinello-Rahman barostat~\cite{pr_barostat} with $2$ ps relaxation time. The rigid body constraints were implemented using the linear constraint solver (LINCS)~\cite{lincs} algorithm. Periodic boundary conditions were applied in all three directions. For the ST2 water, the MD simulations were performed in the $NPT$ ensemble on a system with $N = 2000$ molecules. The electrostatic interactions were truncated at a distance of $0.78$ nm, and the reaction field method was employed to approximate the contribution of electrostatic interactions at longer ranges. The other computational details for this water model are the same as the one employed in Ref.~\cite{st2_sim_2014}.  

We have used different protocols to generate the LDA and HDA structures for these two water models. A schematic diagram of the protocols followed is shown in Fig.~\ref{phase_diagram}. For the TIP4P/2005 water, we first equilibrated liquid water at $T = 240$ K (supercritical to the liquid-liquid critical temperature, $T_c \sim 182$ K) and $P = 1$ bar. The LDA configurations were obtained by isobarically cooling down the equilibrated liquid configurations from $240$ K to $80$ K with a quench rate ($q_T$) of $10$ K/ns. All the LDA configurations were then isothermally compressed at $80$ K from $1$ bar to $3.0$ GPa to get the corresponding HDA structures, named HDA-$l$. Two compression rates were employed for this, $q_P=20$ MPa/ns and $q_P=100$ MPa/ns. We then isothermally decompressed all the HDA-$l$ configurations from $3.0$ GPa to $-0.6$ GPa to get the corresponding low-density configurations, named LDA-$h$. The computational protocol followed to prepare these glassy polymorphs is similar to the one used by Giovambattista and coworkers~\cite{handle2019glass}. To compare the structural features of the LDA with the LDA-$h$, we also isothermally decompressed the LDA configurations to the same $(P, \rho)$ as the LDA-$h$ to get the LDA-II configurations (see Fig.~\ref{rho_e}A). Therefore, the LDA-II configurations have the same ($T,P,\rho$) as the LDA-$h$ configurations, only the protocol of preparation is different. Even though the high-density phase resulting from the isothermal compression of the LDA phase corresponds to the very high-density (VHDA) phase, we adopt the widely used convention of referring to all high-pressure glasses as HDA.   

For the ST2 water, we first equilibrated the liquid water at $T = 230$ K (subcritical to the liquid-liquid critical temperature, $T_c \sim 247$ K) and $P = 1800$ bar to get the thermally equilibrated LDL phase. The LDL configurations were then isobarically cooled at $1800$ bar to $80$ K with a quench rate $q_T=10$ K/ns to get the corresponding LDA structures. HDA-$l$ configurations were then obtained by isothermally compressing the LDA configurations from $1800$ bar to $\sim 2.2$ GPa with compression rates $q_P=20$ MPa/ns and $100$ MPa/ns~\cite{wong2015pressure}. We then isothermally decompressed all the HDA-$l$ configurations from $2.2$ GPa to $-0.5$ GPa to get the corresponding low-density configurations, named LDA-$h$. Similar to the TIP4P/2005 water, we also isothermally decompressed ST2 water's LDA configurations to pressure $P=-0.5$ GPa to get the corresponding LDA-II samples with the same density as of the LDA-$h$ (see Fig.~\ref{rho_e}A). We also obtained the HDA phase from the thermally equilibrated HDL phase at $T = 230$ K and $P = 3200$ bar by rapidly cooling it to $80$ K with $q_T=10$ K/ns. All the results presented in the main text are for $q_T=10$ K/ns and $q_P=20$ MPa/ns. 
\section{\label{sec:level2}Results and Discussion}
\subsection{\label{subsec:21}Global energetic and structural features of the LDA and HDA ices of water}
\begin{figure}
   \centering
    \includegraphics[scale=0.30]{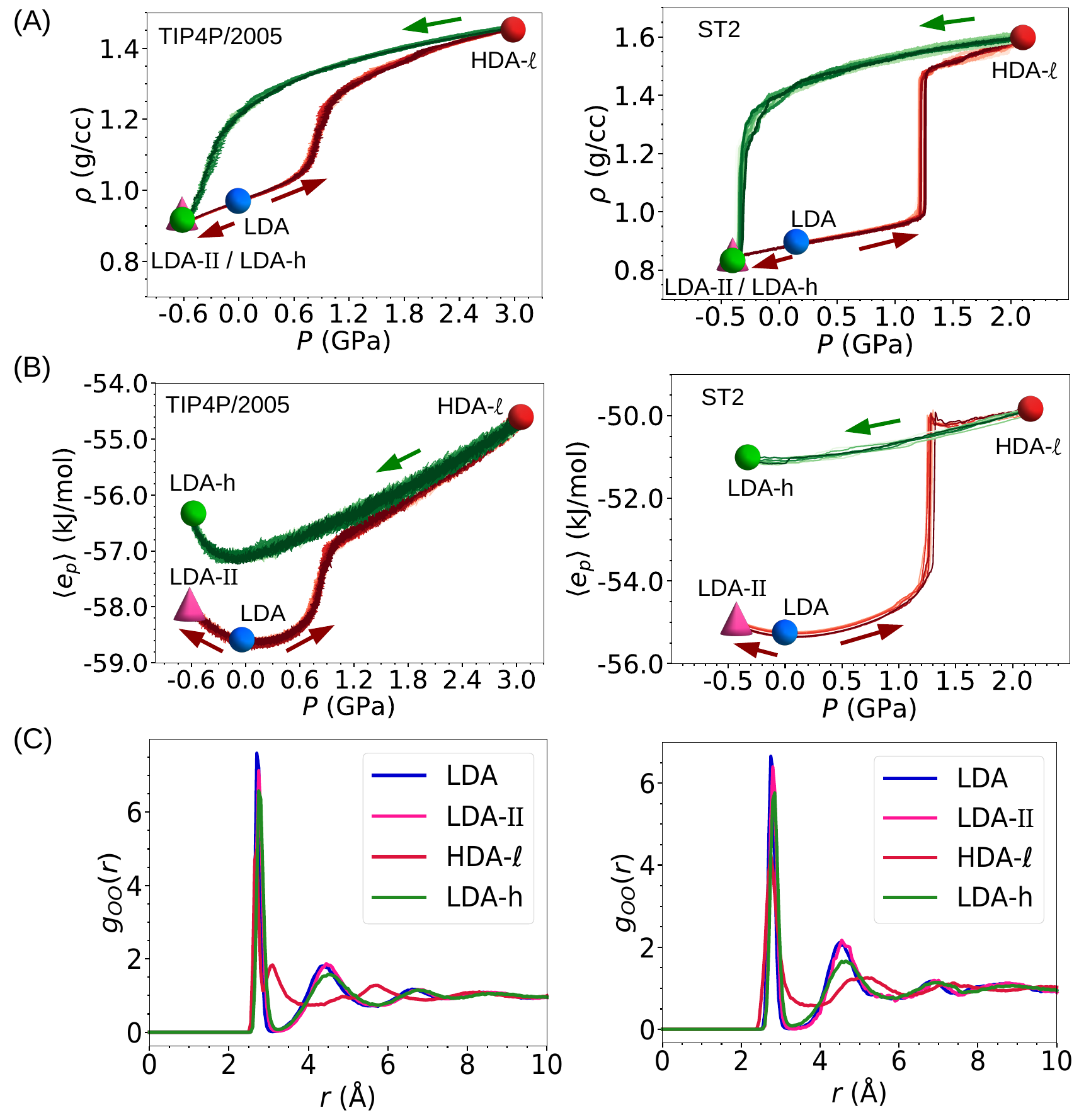}
    \caption{(A) The density ($\rho$) versus pressure ($P$) plot obtained by isothermally compressing ten different LDA samples at $80$ K for TIP4P/2005 and ST2 water. The LDA structures were compressed isothermally with a compression rate, $q_P = 20$ MPa/ns to get the corresponding HDA-$l$ structures (red curve with forward arrow). These HDA-$l$ structures were then decompressed isothermally with the same $q_P$ to get the LDA-$h$ structures (green curve). (B) The average potential energy per particle $\langle e_p \rangle$ versus $P$ along the paths reported in (A) for the TIP4P/2005 and ST2 water. The LDA structures are also isothermally decompressed to get the LDA structures (named as, LDA-II) with the same ($P, \rho$) as the LDA-$h$ (red curve with backward arrow). The LDA, HDA-$l$, and LDA-$h$ are represented by blue, red, and green spheres, respectively. The pink triangle represents the LDA-II phase. (C) The oxygen-oxygen radial distribution function ($g_{OO}(r)$) for the LDA, HDA-$l$, LDA-II, and LDA-$h$ phases for the TIP4P/2005 and ST2 water.}
    \label{rho_e}
\end{figure}

We first explored the global (bulk) structural and energetic features of the LDA and HDA configurations prepared using different protocols outlined in the previous section (see Fig.~\ref{phase_diagram}). As has already been reported in earlier studies~\cite{ nicolas_st2_2013, handle2019glass, engstler2017heating}, during the isothermal compression of LDA to HDA-$l$, the density shows a sharp change as the pressure is increased for both the water models (Fig.~\ref{rho_e}A). We further note that the transition is steeper, and hysteresis is more pronounced for the ST2 than the TIP4P/2005 water. This pronounced hysteresis and steeper transition for the ST2 could be a manifestation of the pressure-induced discontinuous (or, first-order) LDL to HDL transition, unlike the TIP4P/2005 water where the LDL-like phase gradually transitions to HDL-like on compression without encountering any singularity at $T > T_c$. This observation is in agreement with recent seminal work by Gartner et al.~\cite{twostrucutresgartner2021manifestations} which reported the signatures of the LLCP (more precisely, long-range density fluctuations near the LLCP) hidden in the structure of water glasses. The corresponding changes in the average potential energy per particle ($\langle e_{\rm p} \rangle$) of the system are shown in Fig.~\ref{rho_e}B. It is important to note here that the energy difference between the compression and decompression pathways is well separated for both models, even for the LDA-$h$ and LDA-II configurations for which the ($T,P,\rho$) are the same. This suggests that the LDA-II and LDA-$h$ phases are structurally not the same, despite having the exactly same thermodynamic parameters.  

Figure~\ref{rho_e}C shows the oxygen-oxygen radial distribution function ($g_{OO} (r)$) for the different amorphous ice phases. The LDA, LDA-II, and LDA-$h$ exhibit very similar two-body correlations for both the models (Fig.~\ref{rho_e}C)), indicating a high degree of structural similarity with well-separated first and second coordination shells. However, as reported earlier, $\langle e_{\rm p} \rangle$ is significantly different for these two phases, suggesting that $g_{OO}(r)$ fails to capture the structural dissimilarity (if any) in the LDA-II and LDA-$h$ phases. This observation is consistent with both the water models. Therefore, to understand the structural differences responsible for this observed difference in $\langle e_{\rm p} \rangle$ between the two phases, one must probe many-body correlations and medium-range topological features of the amorphous ices. 
\begin{figure*}[ht!]
    \centering
    \includegraphics[scale=0.3]{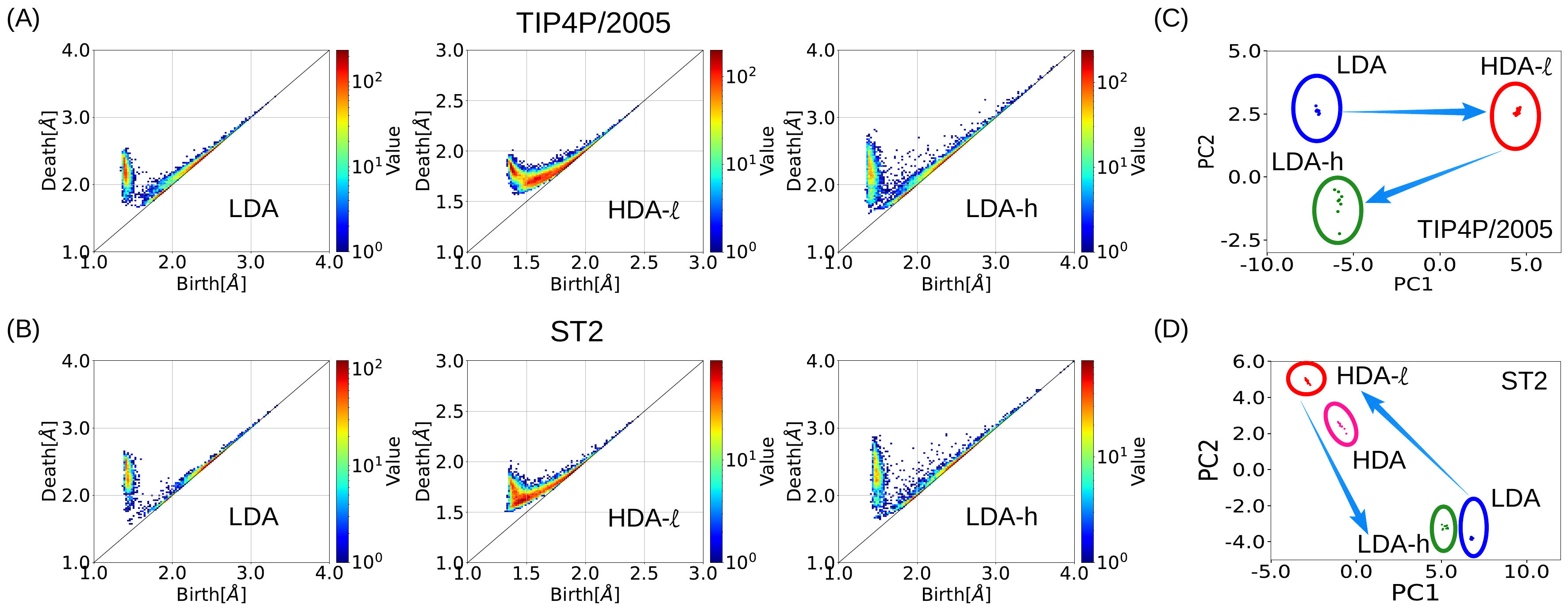}
    \caption{The one-dimensional persistence diagram (PD1) for the LDA, HDA-$l$, and LDA-$h$ phases for the TIP4P/2005 (A) and ST2 (B) water is shown. Here, the birth and death axes are multiplied by the box length to have units in $\AA$. The principal component analysis (PCA) results on the scaled PD1s are shown for the TIP4P/2005 (C) and ST2 (D) water. The LDA, HDA-$l$, and LDA-$h$ configurations are indicated by blue, red, and green circles, respectively. For the ST2 water, we have also included the HDA phase (purple circle) prepared by rapid cooling of the thermally equilibrated HDL phase with $q_T = 10$ K/ns.}
    \label{ph}
\end{figure*}
\subsubsection{Persistent Homology}
Many recent studies suggest that PH is a very effective tool for capturing the structural differences between the amorphous and glassy systems, which otherwise are difficult to quantify using traditional structural order parameters. The details of the PH analysis protocol are briefly described here. Suppose we have an atomic configuration of $N$ atoms with coordinates, $\mathbf{Q}=(\mathbf{x}_1,\mathbf{x}_2,..,\mathbf{x}_N)$ with input radii $\mathbf{R}=(r_1,r_2,...,r_N)$. To compute PH, we place a sphere of radius $r_j$ centered at $x_j$ and increase the radius of the sphere as $\sqrt{r_j^2 + \varepsilon}$ with increasing $\varepsilon$, the scale parameter in the computation. As these spheres intersect, line segments are added to connect corresponding atoms. With the progressive increase of $\varepsilon$, additional line segments emerge, ultimately forming a ring as they connect multiple atoms end-to-end. This is called the birth of a ring, say ring $X$, and the corresponding $\varepsilon$ is recorded as the birth scale $b_X$ of the ring. Further increase in the $\varepsilon$ value results in the expansion of these spheres, leading to their penetration and causing the ring's disappearance. The associated death scale is recorded as $d_X$. Multiple rings may born and subsequently die with changes in the $\varepsilon$ value. The resulting diagram, which plots all the pairs of birth and death for these rings, is called a persistence diagram (PD). Specific $\varepsilon$ values may lead to connections between spheres, creating voids or cavities between them. When a PD is generated solely from ring information, it is denoted as PD1 (one-dimensional PD). Similarly, if it encompasses only cavity information, it is labeled as PD2 (two-dimensional PD). In this work, we have used only PD1 to distinguish the structural differences between the LDA and HDA phases of water.

We have used HomCloud software~\cite{obayashi2022persistent} for the PH analysis. The computed PD1 for our systems is shown in Fig.~\ref{ph}. The PD1 here is computed with only the oxygen atoms with coordinates scaled with the box length (i.e., oxygen atoms are represented by a point cloud inside a cubic box of unit length), referred as ``scaled PD1" in the subsequent sections. Each PD1 consists of three components, the birth scale as the x-axis, the death scale as the y-axis, and the color map representing the density of rings. For the TIP4P/2005 water, the PD1s of the LDA and HDA-$l$ phases exhibit notable differences, indicating distinct global structural and topological features (see Fig.~\ref{ph}A). In the LDA phase, there are extended vertical bands at lower birth scales, implying the presence of larger rings (involving a greater number of atoms) that persist for a wide range of $\varepsilon$ values. The band near the diagonal line consists of smaller rings (involving fewer atoms, mostly three members) with less persistence. We also note that, there is an abundance in the number of rings on the vertical band, suggestive of relatively larger and persistent rings. However, the HDA-$l$ phase shows a merged vertical and diagonal band. The density of the rings is higher in this merged band. This scenario remains the same for the ST2 water as well (see Fig.~\ref{ph}B). Therefore, despite the different protocols followed to generate the HDA phase for the TIP4P/2005 and ST2 water, the topological features show a close resemblance.  

Interestingly, unlike the $g_{OO}(r)$, the PD1 corresponding to the LDA and LDA-$h$ phases exhibit noticeable difference for both the water models. The vertical band in the PD1 of LDA-$h$ appears more pronounced compared to the PD1 of the LDA phase, and the two (vertical and diagonal) bands seem to merge in the LDA-$h$ for both the water models. We employed principal component analysis (PCA) to quantify the structural and topological differences encoded in the scaled PD1 of different amorphous ice phases~\cite{hirata2020structural} (see Section~\ref{pd_details} in the Supplementary Materials). All ten independent simulation results for each of the three phases were utilized in this analysis. Their corresponding scaled PD1s were vectorized, and the entire data set was then projected onto a reduced-dimensional space comprising principal component $1$ (PC1) and principal component $2$ (PC2). The results are shown in Figs.~\ref{ph}C and~\ref{ph}D for TIP4P/2005 and ST2 water, respectively. Again, the underlying topological differences are evident from Figs.~\ref{ph}C and~\ref{ph}D. The LDA and LDA-$h$ phases are distinct but show close structural similarities compared to the HDA phases. 

So far, we have discussed the global structural features of the HDA and LDA phases; however, structural insights at the local level are a prerequisite to understand the microscopic pathway of phase transition. As discussed earlier, compared to the fluid and crystal phases, the identification of an order parameter that can distinguish the local structural environments of different amorphous ice phases is rather a challenging task. In the next section, we have introduced a structural order parameter based on PH that can unambiguously identify the LDA- and HDA-like local environments. In addition, we have also discussed the locally-averaged variant of Steinhardt-Nelson-Ronchetti bond-orientational order parameter (BOP)~\cite{steinhardt1983dr, boattini2020autonomously} and local density as metrics to identify the LDA and HDA-like environments in the system. 
\subsection{\label{subsec:22} Local (microscopic) structural and topological features} 
\subsubsection{Local structural order parameter derived from persistence homology}
\begin{figure} [ht!]
    \centering
    \includegraphics[scale=0.3]{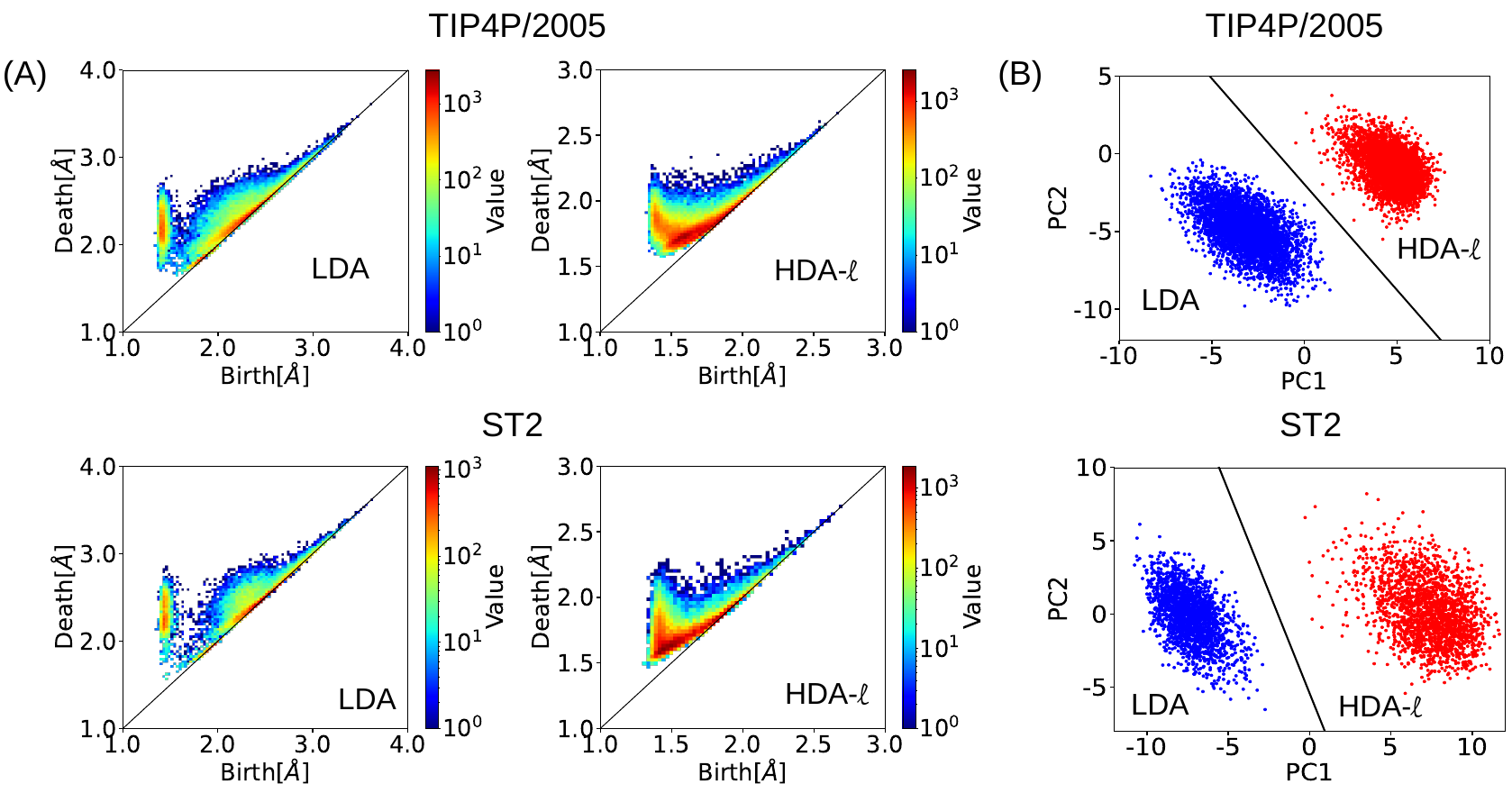}
    \caption{(A) The one-dimensional persistence diagram (PD1) of the system involving only the water's local environment for the LDA and HDA-$l$ phases for the TIP4P/2005 and ST2 water. The local environment around each water molecule is defined by its $20$ neighboring molecules. Here, the birth and death axes are multiplied by the box length to have units in $\AA$. (B) The principle component analysis results: the projection of the scaled PD1s onto the PC1-PC2 plane is shown for both the water models. The solid black line is the Support Vector Machine (SVM) decision boundary that best separates the LDA and HDA-$l$ phases in the PC1-PC2 plane. The directed distance of each projected point in the PC1-PC2 plane from the decision boundary can be used as an order parameter to identify the LDA and HDA-$l$-like local environments.}
    \label{svm}
\end{figure}

To probe the local topological distinction between the LDA and HDA-$l$ phases from the PH analysis, we constructed the particle-wise PD1s~\cite{becker2022unsupervised}. The system is represented by a point cloud consisting of only the oxygen atom coordinates (scaled with the box length) of the water molecules (see Section~\ref{pd_details} in Supplementary Materials for the details). To define the local environment, we considered a central oxygen atom and chose the nearest $n_{\rm p}$ neighbors. We then computed the (scaled) PD1 by involving these $n_{\rm p}+1$ atoms. The PD1 of the system involving only the water's local environment (i.e., involving $n_{\rm p}$ particles around each oxygen atom in the system) for the LDA and HDA-$l$ phases are shown in Fig.~\ref{svm}A. To quantify the structural (topological) differences hidden in the PD1s, we applied the PCA on the vectorized data set of the scaled PD1s and projected the result onto a reduced 2D space of PC1 and PC2. This gives us a scatter plot with $N$ points ($N$ is the number of water molecules in the system). For the compression pathway, the initial LDA and the final HDA-$l$ are projected together on the same PCA plane to get a scatter plot with two distinct point clusters, representing LDA and HDA-$l$ phases (see Fig.~\ref{svm}B). We used the Support Vector Machine (SVM) with linear kernel function~\cite{JMLR:v12:pedregosa11a} to find the equation of the line (called decision boundary) that best separates both these point clusters in the PC1-PC2 plane. Now, the directed distance from this decision boundary to each point in the PC1-PC2 plane can be used as an order parameter to identify the LDA and HDA-$l$-like local environments. This order parameter is named as $\zeta_{\rm p}$. A similar approach is followed for distinguishing the HDA-$l$ to LDA-$h$ phases during the decompression-induced transition.  

We have used $n_{\rm p} = 20$ in this work; however, we have also explored $n_{\rm p}=22$ and $25$ to check the sensitivity of the results on the choice $n_{\rm p}$ and found that results are not remarkably sensitive to $n_{\rm p}$. The number $n_{\rm p}=20$ or higher indicates that we are well beyond the first shell and, hence, exploring the local topological features at medium--range scales. The distribution of $\zeta_{\rm p}$ with $n_{\rm p}=20$ is shown in Fig.~\ref{op3}A and Fig.~\ref{op3}D for the TIP4P/2005 and ST2 water, respectively. It is evident that, for both the models, $\zeta_{\rm p}$ unambiguously differentiates between LDA and HDA-$l$-like local environments.     
\subsubsection{Locally-averaged bond-orientational order parameter}
\begin{figure}
    \centering
    \includegraphics[scale=0.26]{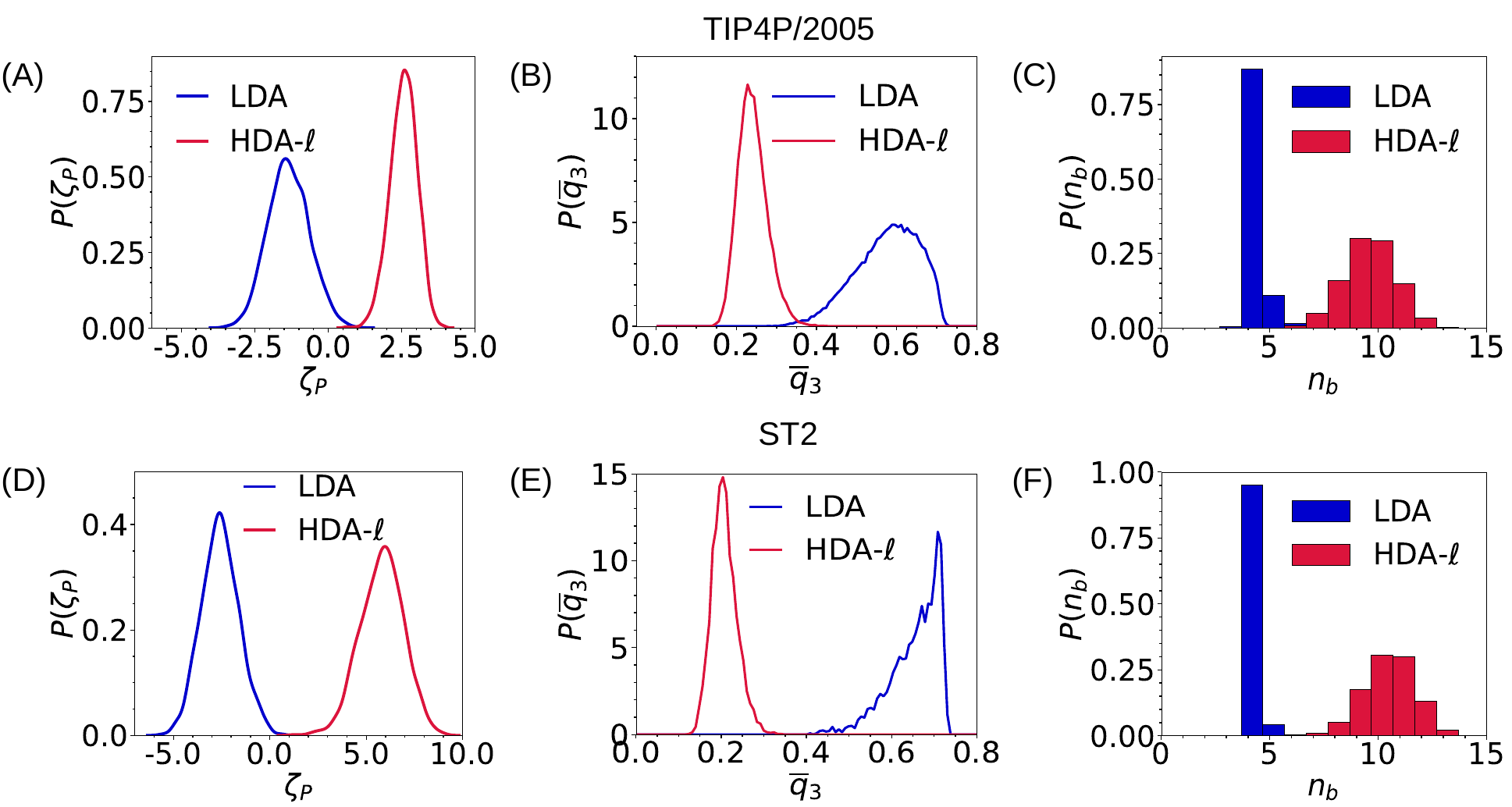}
    \caption{The local order parameter ($\zeta_{\rm p}$, $\overline{q}_3$, and $n_{\rm b}$) distributions for the LDA (blue) and HDA-$l$ (red) phases for the TIP4P/2005 (top) and ST2 (bottom) water.}
    \label{op3}
\end{figure}

Locally-averaged BOPs provide a robust way to explore the structural differences between different condensed (crystalline and liquid) phases~\cite{tanaka_rev_bo}. Here, we have used the coarse-graining approach proposed by Boattini et al.~\cite{boattini2020autonomously} to define the order parameter to distinguish the local structural features of the LDA and HDA-$l$ phases. The locally-averaged BOP ($\overline{q}_l$) is given as:
\begin{equation}
    q_{lm} = \frac{1}{n_b(i)} \sum_{j \in n_b(i)} Y_l^m(\mathbf{r}_{ij})
\end{equation}
\begin{equation}
    q_{l}(i) = \sqrt{\frac{4 \pi}{2l+1} \sum_{m=-l}^{l} |q_{lm}(i)|^2}
\end{equation}
\begin{equation}
    \overline{q_{l}}(i) = \frac{1}{n_b(i)+1} \left[q_l(i) + \sum_{k \in n_b(i)} q_l(k)\right].
\end{equation}
Here, the functions $Y_l^m$ are the spherical harmonics, and $n_b(i)$ is the number of nearest neighbors of particle $i$. The advantage of using the coarse-grained version of the order parameter is that it can include the structural features up to the second (or, beyond) coordination shell. We must note that the cutoff value for calculating the number of nearest neighbors should be carefully chosen. Since the two phases differ significantly in density, a common distance cutoff would incorporate more particles in local averaging for the HDA phase compared to the LDA phase. Therefore, the separation between the $\overline{q}_l$ distributions corresponding to the LDA and HDA phases might not be solely due to the underlying local structural differences but could also be due to averaging over more particles for the HDA phase. To minimize the over-coarse-graining for the HDA-$l$ phase, we scaled the distance $r$ with inter-particle separation (or, density) as, $\zeta \equiv r\rho^{1/3}$, and chose a unique cutoff distance from the scaled radial distribution function $g_{\rm OO}(\zeta)$ (see Fig.~\ref{rescaled_rdf} in the Supplementary Materials). We have used the scaled cutoff distance, $\zeta_c = 1.20$ to define the neighbor list for the LDA and HDA-$l$ phases. After this, we calculated the $\overline{q}_l$ distributions with $l = 1,..,12$ for the LDA, and HDA-$l$ phases, and found that $\overline{q}_3$ satisfactorily distinguishes the LDA and HDA-like local environments for both the water models (see Figs.~{\ref{op3}B and~\ref{op3}E}). The $\overline{q}_3$ distribution for the HDA-$l$ and LDA-$h$ phases is shown in Fig.~\ref{ql_dist} of the Supplementary Materials. We have also checked the sensitivity of the results on the choice of the scaled cutoff distance by using lower ($\zeta_c = 1.15$) and higher ($\zeta_c = 1.25$) cutoff distances and found that the results are not delicately sensitive to the choice of $\zeta_c$ (see Fig.~\ref{q3_cutoff} in the Supplementary Materials).  
\subsubsection{Local density}
As the LDA and HDA-$l$ phases have significantly different densities, the local density is a natural choice for an order parameter to assign the LDA and HDA-like local environments. This order parameter has been used recently to probe the pathways of LDA to HDA-$l$ transition~\cite{nicolas_st2_2013, vmolineroFOPT}. The local density around a central water molecule can be defined in terms of the number of neighboring molecules ($n_{\rm b}$) within an oxygen-oxygen radial cutoff distance of $r_c$ ($3.5~\AA$ in this work) from the tagged central water molecule. The $n_{\rm b}$ distribution for the LDA and HDA-$l$ phases are shown in Fig.~\ref{op3}C for the TIP4P/2005 and in Fig.~\ref{op3}F for the ST2 water. It is evident from the figures that, for both the models, $n_{\rm b}$ can distinguish between LDA and HDA-$l$-like local environments. However, we must note here that the local density encodes only the local packing information, not the local structural and topological features hidden in the amorphous ice.  
\begin{figure*}
    \centering
    \includegraphics[scale=0.25]{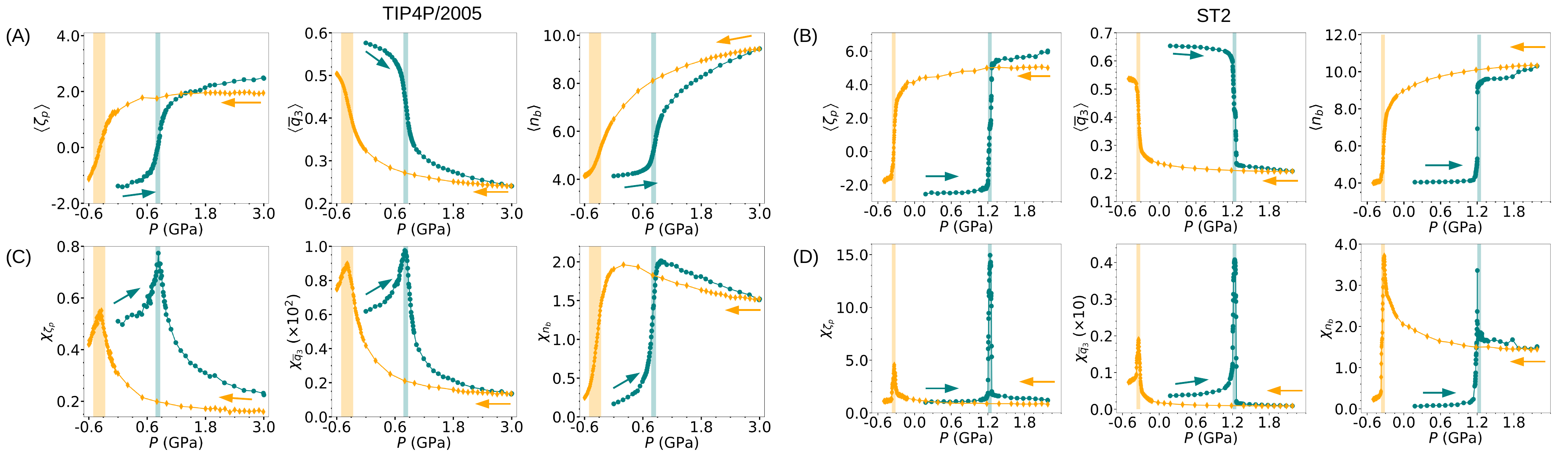}
    \caption{The evolution of the average order parameters with pressure ($P$) on isothermal compression (teal color) and decompression (orange color) of the system at $80$ K for the TIP4P/2005 (A) and ST2 (B) water is shown. The transition for the ST2 water is much more drastic compared to the TIP4P/2005 water. The local order parameter susceptibility, $\chi_{\rm OP}$ (with OP = $\zeta_{\rm p}$, $\overline{q}_3$, and $n_{\rm b}$) change on compression (teal color) and decompression (orange color) at $80$ K for the TIP4P/2005 (C) and ST2 (D) water. The shaded regions are bounded by the lowest and highest $P^*$ values obtained from the ten independent simulations for each compression and decompression-induced transition.}
    \label{opavg}
\end{figure*}
\begin{figure}
    \centering
    \includegraphics[scale=0.68]{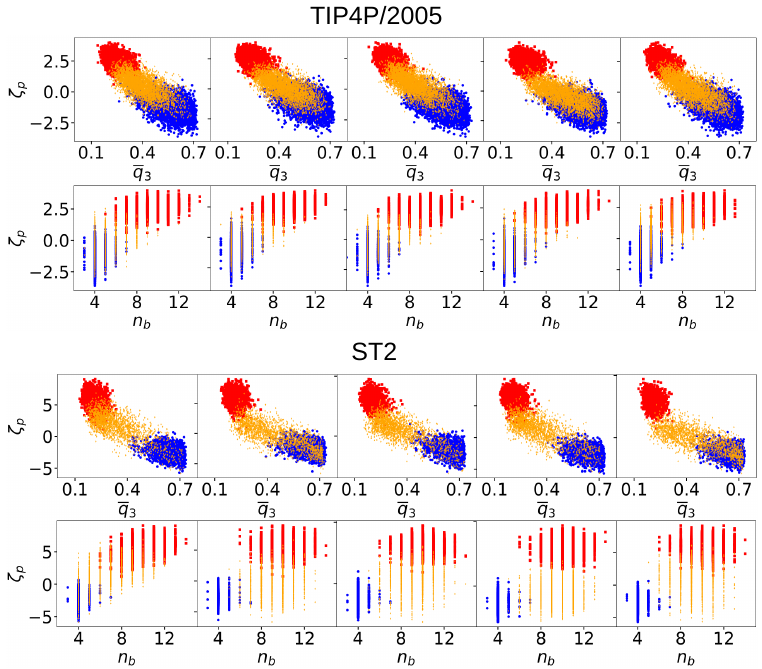}
    \caption{A scatter plot showing the projection of the LDA (blue), HDA-$l$ (red), and the system near the transition pressure, $P^*$ (orange) in $\overline{q}_3 - \zeta_{\rm p}$ and $n_{\rm b} - \zeta_{\rm p}$ planes for the TIP4P/2005 (top) and ST2 (bottom) water. We have shown the scatter plot for $5$ independent trajectories. We note enhanced order parameter fluctuations near $P^*$ where along with the LDL and HDL-like local environments, particles with intermediate structural order (named as, intermediate phase (IP) particles) are also present.}
    \label{sc_plot}
\end{figure}
\subsection{\label{subsec:23} Evolution of the local order parameter during the transition between the amorphous ices}
In order to gain deeper insights into the nature of the pressure-induced transition between the two amorphous ice phases, we have examined the evolution of the average local structural order parameter during the transition, especially near the transition pressure, $P^*$. $P^*$ is defined as the pressure corresponding to the maximum of the derivative of the density vs. pressure plot (Fig.~\ref{rho_e}A). In Figs.~\ref{opavg}A and~\ref{opavg}B, we report the evolution of the average local order parameters on compression and decompression for both the water models. It can be observed that the average value of the order parameters rapidly changes in the close vicinity of $P^*$ indicating a structural phase transition. Note that, for the HDA-$l$, $\left \langle n_{\rm b} \right \rangle$ and $\left \langle \overline{q}_3 \right \rangle$, values overlap exactly, and so the decompression curve starts exactly from where the compression curve ends. However, for $\left \langle \zeta_{\rm p} \right \rangle$, this alignment is not always guaranteed due to the nature of the definition of this order parameter. $\zeta_{\rm p}$ was defined as the directed distance from the decision boundary that best separates the LDA (LDA-$h$) and HDA-$l$ phases on compression (decompression). Hence, the value of $\zeta_{\rm p}$ for HDA-$l$ can differ as the decision boundary is not the same for the compression and decompression trajectories even though the HDA-$l$ phase under consideration is the same. We further note that, the ST2 model (where the LDA phase prepared by quenching the LDL phase at subcritical temperature, $T < T_c$) shows sharper order parameter changes, compared to the TIP4P/2005 where the LDA phase is generated by quenching the LDL-like water at supercritical temperature ($T > T_c$). This sharper LDA to HDA-$l$ transition in the order parameter space and more pronounced hysteresis on reversal for the ST2 (compared to the TIP4P/2005 water) water could be a manifestation of the first-order LDL to HDL transition at subcritical temperatures. 

\begin{figure*} [ht!]
    \centering
    \includegraphics[scale=0.7]{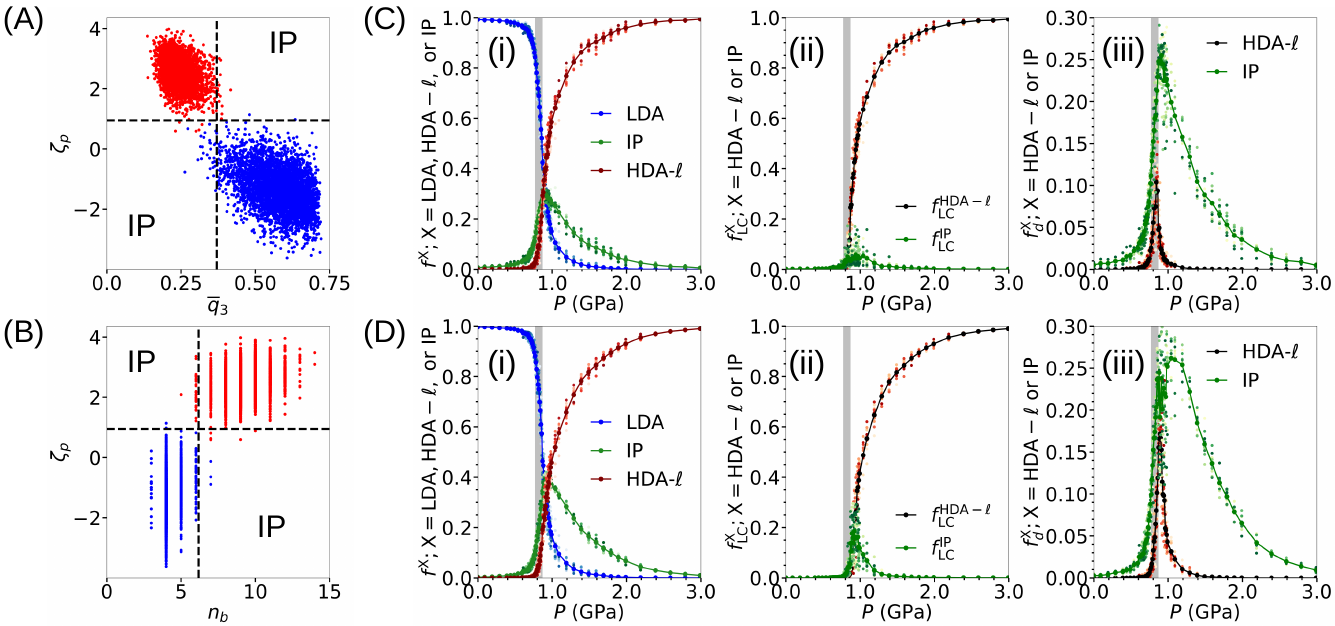}
    \caption{The LDA to HDA-$l$ phase transition pathway in the cluster-size space for the TIP4P/2005 water. The LDA, HDA-$l$, and pre-ordered intermediate phase (IP)-like local environments are identified using a combination of two order parameters: $\zeta_{\rm p}$ and $\overline{q}_3$ (A); and $\zeta_{\rm p}$ and $n_{\rm b}$ (B). The blue and red scatter points represent the LDA and HDA-$l$ phases, respectively. The dashed lines represent the order parameter boundaries that best separate the LDA and HDA-$l$ phase distributions along the respective order parameter. The clustering analysis results on structural inhomogeneities classified using (A) and (B) are shown in (C) and (D), respectively. The $P$-dependent fraction of (i) the LDA-like, IP, and HDA-$l$-like particles, (ii) the HDA-$l$-like and IP particles that are part of the respective largest cluster ($f_{\rm LC}^{\rm X}$, with X = HDA-$l$ or IP), and (iii) the HDA-$l$ and IP particles that are not part of the respective largest cluster ($f_d^{\rm X}$, with X = HDA-$l$ or IP) is reported. The scatter points represent results from all $10$ independent trajectories and the solid lines indicate the average behavior. The shaded region is bounded by the smallest and largest $P^*$ values evaluated for each independent compression trajectory of the LDA phase.}
    \label{clustf}
\end{figure*}

To quantify the nature of the local structural fluctuations, in Figs.~\ref{opavg}C and~\ref{opavg}D, we report the local order parameter susceptibility $\chi_{\rm OP}$, defined as, $\chi_{\rm OP} = \langle \rm OP^2 \rangle - \langle \rm OP \rangle^2$, where $\rm OP$ corresponds to $\zeta_{\rm p}$, $\overline{q}_3$, and $n_{\rm b}$; and $\langle ... \rangle$ indicates average over the number of particles. The maximum in $\chi_{\rm OP}$ for all the three order parameters suggests enhanced local structural heterogeneities near $P^*$. To visualize these enhanced local order parameter fluctuations, in Fig.~\ref{sc_plot}, we show the two-dimensional projection of the system on $\overline{q}_3 - \zeta_{\rm p}$ and $n_{\rm b} - \zeta_{\rm p}$ planes for both the water models. As evident from the figure, at $P^*$, some particles are LDA-like, some are HDA-$l$-like and many particles don't fall into either of these two categories. These particles, which are neither LDA-like nor HDA-$l$-like, are assigned as pre-ordered intermediate phase (IP) particles (see also Figs.~\ref{clustf}A and~\ref{clustf}B). Upon compression, the system shifts from the LDA region to the HDA-$l$ region in a rather continuous fashion (see also the local OP distributions shown in Fig.~\ref{opdist} of the Supplementary Materials). As a result, the number of IP particles increases first and then gradually decreases. Only the compression trajectory is represented here, however, the decompression trajectory also shows a similar transition mechanism. The results shown above suggest that the compression-induced LDA to HDA-$l$ transition occurs via a pre-ordered intermediate phase. 
\begin{figure}
    \centering
    \includegraphics[scale=0.45]{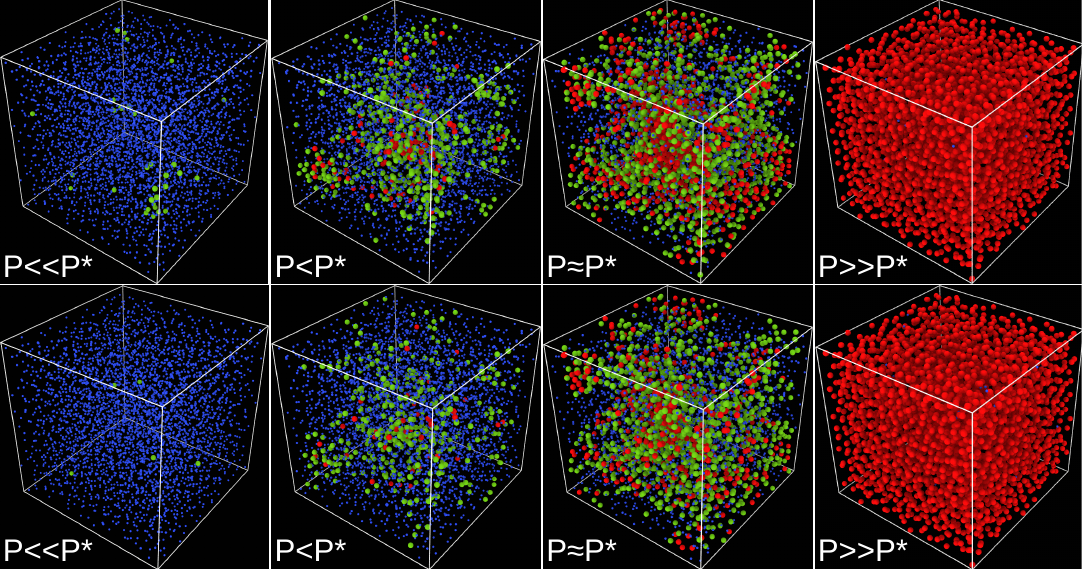}
    \caption{Representative snapshots of the TIP4P/2005 water system at different pressures along isothermal compression-induced LDA to HDA-$l$ transition are shown. The LDA-like, IP, and HDA-like particles are represented by blue, green, and red spheres, respectively. The different types of local environments in the system are identified using a combination of $\zeta_{\rm p}$ and $\overline{q}_3$ (top), and $\zeta_{\rm p}$ and $n_{\rm b}$ (bottom) order parameters (see Figs.~\ref{clustf}A and~\ref{clustf}B).}
    \label{snap}
\end{figure}
\subsection{\label{subsec:24}LDA to HDA-$l$ transition pathway in the cluster-size space}
To gain further deeper insights into the nature of transition, we performed clustering analysis to probe the spatial distribution of the different types of local environments in the system and its dependence on $P$. We first broadly classified the different types of local environments into three categories ---  LDA, HDA-$l$, and IP-like. The different types of local environments are identified using a combination of two order parameters: $\zeta_{\rm p}$ and $\overline{q}_3$ (Fig.~\ref{clustf}A); and $\zeta_{\rm p}$ and $n_{\rm b}$ (Fig.~\ref{clustf}B). The particles that neither belong to HDA-$l$ nor the LDA are classified as IP-like. We note that a binary classification involving only the new and old phases is often employed to characterize the pathways of phase transition. This binary classification, however, might not give us a deeper understanding of the transition mechanism, especially when a significant number of pre-ordered IP particles are present in the system near $P^*$ (see Fig.~\ref{sc_plot}). This aspect was recently studied by Garkul and Stegailov~\cite{garkul2022molecular} where using local density as an order parameter, they divided the neighbor range into three groups: atoms with fewer neighbors (LDA-like), atoms with more neighbors (HDA-$l$), and intermediate neighbor atoms. They excluded these intermediate neighbor/density particles from the analysis. However, an in-depth examination of the role of these intermediate-order particles is desirable to gain a comprehensive understanding of the transition mechanism.    

To study how the HDA-$l$ phase appears in the background LDA phase on compression, we calculated the following three quantities --- (i) the fraction of the new phase $f^{\rm X}$, (ii) the fraction of $\rm X$ that is part of the largest cluster ($f_{\rm LC}^{\rm X}$), and (iii) the fraction of $\rm X$ that is not part of the largest cluster ($f_d^{\rm X}$), $f_d^{\rm X} = f^{\rm X} - f^{\rm X}_{\rm LC}$, in the system (see Figs.~\ref{clustf}C and~\ref{clustf}D for the TIP4P/2005 water). Here, $\rm X$ represents either the growing IP or the HDA-$l$ phase. The largest cluster of the growing HDA-$l$ or IP phase is defined as the biggest collection of HDA-$l$/IP-like particles sharing a common neighborhood defined by a radial cutoff distance of $r_{\rm c} = 3.0~\AA$ (we have also explored a higher, $3.2~\AA$, cutoff distance for this analysis to check the cutoff distance dependence of the results and found that results are not delicately sensitive to the choice of $r_{\rm c}$). Monitoring the size of the largest cluster $f_{\rm LC}^{\rm X}$ allows to probe the localization/connectivity of the structural heterogeneities in the system. A high value of $f_d^{\rm X}$ indicates that the majority of the $\rm X$ phase particles are delocalized inside the system.    
\begin{figure*} [htbp!]
    \centering
    \includegraphics[scale=0.36]{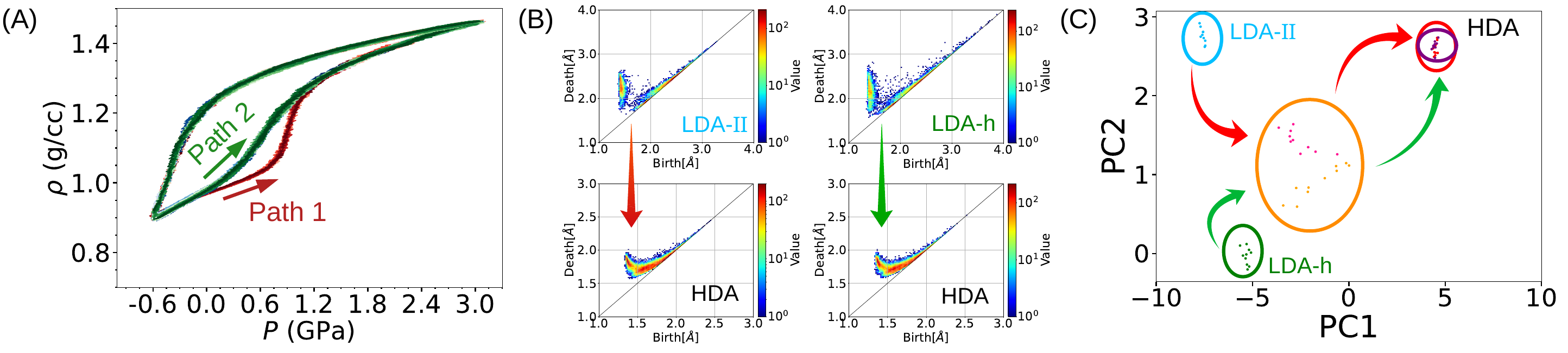}
    \caption{(A) The isothermal compression and decompression paths for the LDA-II (red, Path 1) and LDA-$h$ (green, Path 2) are shown for the TIP4P/2005 water in the $P-\rho$ plane at $80$ K. Note that the second compression and decompression cycle path overlaps completely with the LDA-$h$ compression and decompression path. (B) The persistence diagram of the LDA-II and LDA-$h$ phases along with their corresponding HDA phases. (C) The results of the PCA analysis are shown. We note that, despite the LDA-II and LDA-$h$ showing distinct structural features, the structural features of the final HDA phases are the same. The orange circled region indicates the projection of the configurations near the respective transition pressure $P^*$ for Path 1 (pink dots) and Path 2 (orange dots) on the PC1-PC2 plane.}
    \label{pd_memory}
\end{figure*} 

At lower pressures $(P<<P^*)$, the system has nearly zero HDA-$l$-like and IP particles. On compression, the IP particle population first grows followed by the HDA-$l$-like particles' population (Figs.~\ref{clustf}C(i) and~\ref{clustf}D(i)). Notably, we observe that the IP particle fraction in the system shows a maximum in the vicinity of $P^*$ for both the order parameter criteria used to identify the IP particles. We further observe that very few IP particles are part of the largest cluster, and the majority of the particles are delocalized inside the system (Figs.~\ref{clustf}C(ii \& iii) and~\ref{clustf}D(i \& iii)) --- suggesting a weak (spatial) correlation among the IP particles. Upon further compression, many IP particles get converted into the HDA-$l$ giving rise to a decrease in the IP particle fraction, and in turn, a rapid increase in the fraction of HDA-$l$-like particles in the system. We also note that the $f_d^{{\rm HDA}\mbox{-}l}$ decreases much faster compared to the $f_d^{\rm IP}$ above the transition pressure $P^*$. This slower decrease suggests that the IP particles/clusters are distributed randomly inside a system-spanning network of HDA-like particles and eventually get converted to the HDA-$l$ phase leading to a complete phase transition. These IP particles may show close structural resemblance with the recently discovered medium-density amorphous (MDA) ice phase~\cite{mda_sci,giovam_2024_comm_chem,fausto_mda}. The analysis performed on the ST2 water shows a similar mechanism, but a much sharper transition (see Fig.~\ref{clust_st2} in the Supplementary Materials).        

The above observations point towards a scenario where most LDA particles transition to the HDA-$l$ via a pre-ordered IP-like local environment (LDA $\rightarrow$ IP $\rightarrow$ HDA-$l$). The representative snapshots of the system supporting the above mechanism are shown in Fig.~\ref{snap} (see Fig.~\ref{snaps_st2} in the Supplementary Materials for the ST2 water). Near $P^*$, we can see several clusters of HDA-like particles along with the IP particle clusters are distributed almost evenly throughout the system. We also observe that the shapes of the growing HDA-$l$ clusters are highly ramified. When the IP particles are not identified (for the case of binary phase classification), the LDA phase undergoes a direct transition to the HDA-$l$ via a spinodal-decomposition-like mechanism (see Fig.~\ref{rhol_clust_binary} in the Supplementary Materials), as is also suggested in recent computational studies~\cite{nicolas_st2_2013, mollica2022decompression,vmolineroFOPT}. 

To sum, the LDA to HDA transition mechanism observed here does not seem to suggest a first-order-like transition following nucleation and subsequent growth (in the spirit of the classical nucleation theory) for both the water models; unlike silicon, where Fan et al. recently reported a first-order LDA to HDA transition following nucleation and growth mechanism~\cite{tanaka_natcom_2024}. We also note here that, unlike the equilibrium (crystalline) solid-solid collapse transition, our results do not show any signatures of the (non-classical) wetting-mediated transition mechanisms where the growing final collapsed HDA-$l$ phase is wetted by intermediate-order (or, IP) particles to minimize the surface energy~\cite{singh_jcp_2013, singh_jpcb_2013}.      
\subsection{\label{subsec:25} LDA to HDA transition pathway: Memory effects}
Glasses are non-ergodic systems. Therefore, it is expected that the pathway of transition between the two glassy polymorphs of water may depend on the protocol followed to prepare the initial phase~\cite{wong2015pressure,PELgiovambattista2017influence, engstler2017heating}. That is, the microscopic transition pathway in the structural order parameter space may contain the memory of the protocol followed to get the initial amorphous (glassy) phase. To probe this, we used two samples of LDA --- LDA-$h$ and LDA-II --- with the same thermodynamic parameters, $\rho$, $P$, and $T$ (see Fig.~\ref{rho_e}). We isothermally compressed both the phases with a compression rate $q_P=100$ MPa/ns and decompressed the corresponding HDA phases with the same rate. The density change with pressure along these paths is shown in Fig.~\ref{pd_memory}A. We observe that the LDA to HDA transition paths corresponding to the LDA-II and LDA-$h$ phases considerably differ in the $P-\rho$ plane at intermediate pressures, eventually converging at higher pressure. The LDA-II phase shows a sharper transition compared to the LDA-$h$ phase. We also explored the $P$-dependence of the different local structural order parameters ($\zeta_{\rm p}$, $\overline{q}_3$, and $ n_{\rm b}$) and their susceptibilities and found that they also show a similar sharper transition for the LDA-II phase (see Fig.~\ref{avg_dist_sus_memory} in the Supplementary Materials). We also observe from  these figures that the transition pressure for the LDA-$h$ is at a lower $P$ compared to the LDA-II phase.  
\begin{figure}
    \centering
    \includegraphics[scale=0.315]{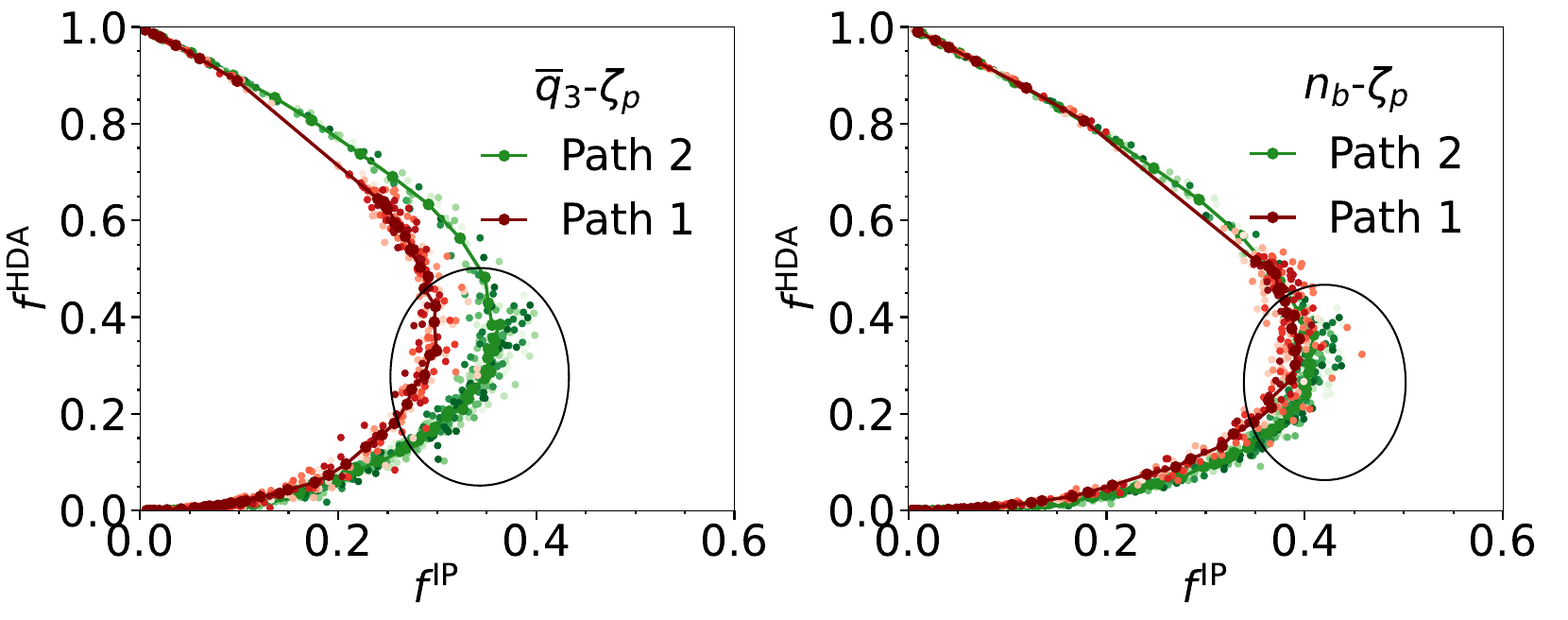}
    \caption{The compression-induced LDA to HDA transition path in the $f^{\rm IP} - f^{\rm HDA}$ plane for the LDA-II (Path 1) and LDA-$h$ (Path 2) phases is shown. Here, $f^{\rm IP}$ indicates the fraction of the IP particles, and $f^{\rm HDA}$ is the fraction of the HDA-like particles in the system. The LDA-like, IP, and HDA-like particles are identified using a combination of two order parameters: $\overline{q}_3$ and $\zeta_{\rm p}$ (left); and $n_{\rm b}$ and $\zeta_{\rm p}$ (right). The open circles represent the fraction of HDA-like and IP particles near the transition pressure $P^*$.}
    \label{path}
\end{figure}

The structural analysis of the LDA (both LDA-$h$ and LDA-II) and the corresponding HDA phases was done using the PH method. The PD1 of the LDA-$h$ and LDA-II phases are shown in Fig.~\ref{pd_memory}B along with the corresponding HDA phases. The PCA results of the scaled PD1s are shown in Fig.~\ref{pd_memory}C. The final HDA phases show very similar structural features indicating that the final HDA phase obtained after compression does not retain the memory of how the initial (parent) LDA phase was prepared --- only the transition path retains this memory. We further studied the detailed transition mechanism by probing the evolution of the fraction of the IP ($f^{\rm IP}$) and the HDA-like ($f^{\rm HDA}$) particles in the system during the LDA to HDA transition. We followed the same procedure outlined in the previous section to identify the LDA, IP, and HDA-like particles in the system (see Figs.~\ref{cluster_memory}A and~\ref{cluster_memory}B in the Supplementary Materials). In Fig.~\ref{path} we report the transition path in the $f^{\rm IP} - f^{\rm HDA}$ plane for the LDA-II (Path 1) and LDA-$h$ (Path 2) phases (the detailed clustering analysis results are shown in Fig.~\ref{cluster_memory}C and~\ref{cluster_memory}D of the Supplementary Materials). We found again a two-step non-diffusive collective transition via a pre-ordered intermediate phase for both the LDA phases (Path 1 and 2). The $f^{\rm IP}$ reaches a maximum in the vicinity of $P^*$ and decreases on further increasing the pressure of the system (see Fig.~\ref{path}). We also observe that the $f^{\rm IP}$ in the system near $P^*$ is slightly higher for Path 2 compared to Path 1. Therefore, despite tracing different paths in the order parameter space, the (microscopic) transition mechanism remains qualitatively the same for both the LDA phases.  
\section{\label{sec:c}Conclusions}
In this work, we have employed computer simulations to probe the microscopic pathway of pressure-induced non-equilibrium LDA to HDA transition. An order parameter capable of unambiguously distinguishing local LDA and HDA-like environments is a prerequisite for this study. Using PH and machine learning, we introduced a new order parameter $\zeta_{\rm p}$ that unambiguously identifies the LDA and HDA-like local environments. We have also used other choices of order parameters, such as locally-averaged BOP and local density, to probe the sensitivity of the transition pathway on the choice of the order parameter. We report that the system transitions continuously and collectively in the order parameter space via a pre-ordered intermediate phase during the isothermal compression and decompression. The local order parameter susceptibility shows a maximum in the vicinity of $P^*$ --- suggesting that structural heterogeneities reach a maximum near $P^*$. The LDA phase prepared by rapidly cooling the LDL at subcritical temperature ($T < T_c$) shows sharper order parameter changes with more pronounced hysteresis on reversal compared to the LDA phase generated by quenching the LDL-like liquid water at supercritical temperature ($T > T_c$). This observation suggests that the sharper LDA to HDA transition for the former could be a manifestation of the first-order LDL to HDL transition.   

To gain deeper insights into the transition pathway, we probed the spatial distributions of the new (pre-ordered intermediate phase and HDA-like) particles/clusters inside the parent LDA phase, and their evolution with the pressure of the system. This analysis unambiguously suggested a two-step non-diffusive LDA to HDA transition mediated via pre-ordered intermediate phase particles. The IP and HDA-like particles/clusters are spatially delocalized inside the LDA phase near $P^*$. We also found that the shapes of the growing HDA-like clusters are highly ramified --- suggesting that surface free energy does not play an important role in deciding the shape and morphology of the growing clusters, unlike the equilibrium diffusive systems. We further investigated the (geometrical) structures and topologies of the LDA and HDA ices formed via different protocols. Unlike the radial distribution functions, the PH-based method is able to capture the structural differences of these different glassy polymorphs prepared via different routes. In addition, the LDA to HDA transition pathways in order parameter space also carry a memory of the protocol followed to prepare the initial LDA phase. 

On a more general note, the PH-based method employed here is not restricted to the system under consideration, and provides a robust way of probing (microscopic) phase transition pathways between any two condensed (equilibrium and non-equilibrium) phases. Recently, many tetrahedral network-forming liquids have been reported to show water-like rich liquid polymorphism (for example, sulfur~\cite{s_llps, s_llps}, phosphorous~\cite{p_llps}, silicon~\cite{si_llps}, tetrahedral network-forming colloidal fluids~\cite{sciortino_natphys_1, sciortino_natphys_2}), including the monoatomic models~\cite{xu2011waterlike, monoatomic_glass_2014, mono_2022, mono_2018}. It would be an interesting avenue for future research to probe the microscopic pathways of LDA to HDA transition and its possible connection with the bulk liquid polymorphism and (anomalous) thermodynamic behavior of these systems. Our approach can also be used to gain deeper structural and topological insights into the recently discovered MDA ice phase of water~\cite{mda_sci}. 
 
\begin{acknowledgments}
R.S.S. thanks Jeremy C. Palmer for providing the molecular dynamics simulation code for ST2 water. R.S.S. acknowledges financial support from DST-SERB (Grant No. SRG/2020/001415 and CRG/2023/002975) and Indian Institute of Science Education and Research (IISER) Tirupati. D. K. acknowledges financial support from DST-SERB (Grant No. CRG/2023/002975). G. R. acknowledges financial support from IISER Tirupati. V.M. thanks DST-INSPIRE for the financial support. The computations were performed at the IISER Tirupati computing facility and at PARAM Brahma at IISER Pune. 
\end{acknowledgments}

%%%%%%%%%%%%%%%%%%%%%%
\bibliographystyle{apsrev4-2}
\bibliography{ref}
\balance
%%%%%%%%%% Merge with supplemental materials %%%%%%%%%%
\pagebreak
\widetext
\begin{center}
\textbf{\large Supplementary Materials}
\end{center}
%%%%%%%%%% Merge with supplemental materials %%%%%%%%%%
%%%%%%%%%% Prefix a "S" to all equations, figures, tables and reset the counter %%%%%%%%%%
\setcounter{equation}{0}
\setcounter{figure}{0}
\setcounter{table}{0}
\setcounter{page}{1}
\setcounter{section}{0}
\makeatletter
\renewcommand{\theequation}{S\arabic{equation}}
\renewcommand{\thefigure}{S\arabic{figure}}
%\renewcommand{\bibnumfmt}[1]{[S#1]}
%\renewcommand{\citenumfont}[1]{S#1}
%%%%%%%%%% Prefix a "S" to all equations, figures, tables and reset the counter %%%%%%%%%%

\section{Persistence diagram (PD) analysis}\label{pd_details}
We have used only the oxygen atoms for the computation of the persistence diagrams. We first scaled the oxygen atom coordinates with the corresponding system's box length to generate a new configuration, called a point cloud. We used alpha filtration to construct all the PDs [see the Ref: Obayashi et al. J. Phys. Soc. Jpn, 91, 091013 (2022)]. We divided the PDs into $128 \times 128$ grids, and after this, each PD was vectorized by the persistence image technique [Ref: Adams et al., J. Mach. Learn. Res. 1, 18 (2017)]. Vectorization converts each PD into a 1D array of $128 \times 128 = 16,384$ elements. Hence, $n$ number of PDs will give a 2D array of dimensions [$n \times 16,384$]. This matrix is provided as an input to the PCA to reduce the dimension to $n \times 2$. For example, in the case of global PDs, we have vectorized all the PDs of $10$ independent simulations for each phase and provided them as an input for the PCA. This PCA reduces the dimension of the dataset from $30 \times 16,384$ ($10$ PDs for each phase and $3$ such phases) to $30 \times 2$. However, in the case of the local PDs, we have $N$ such PDs for each phase, where $N$ is the system size. The PCA was performed on this $2N \times 16,384$ ($N$ for each phase, and $2$ such phases) data and reduced it to $2N \times 2$. Note that, this work focuses on the one-dimensional homology, or PD1, which only has information about rings.

\begin{figure} [!ht]
    \centering
    \includegraphics[scale=0.5]{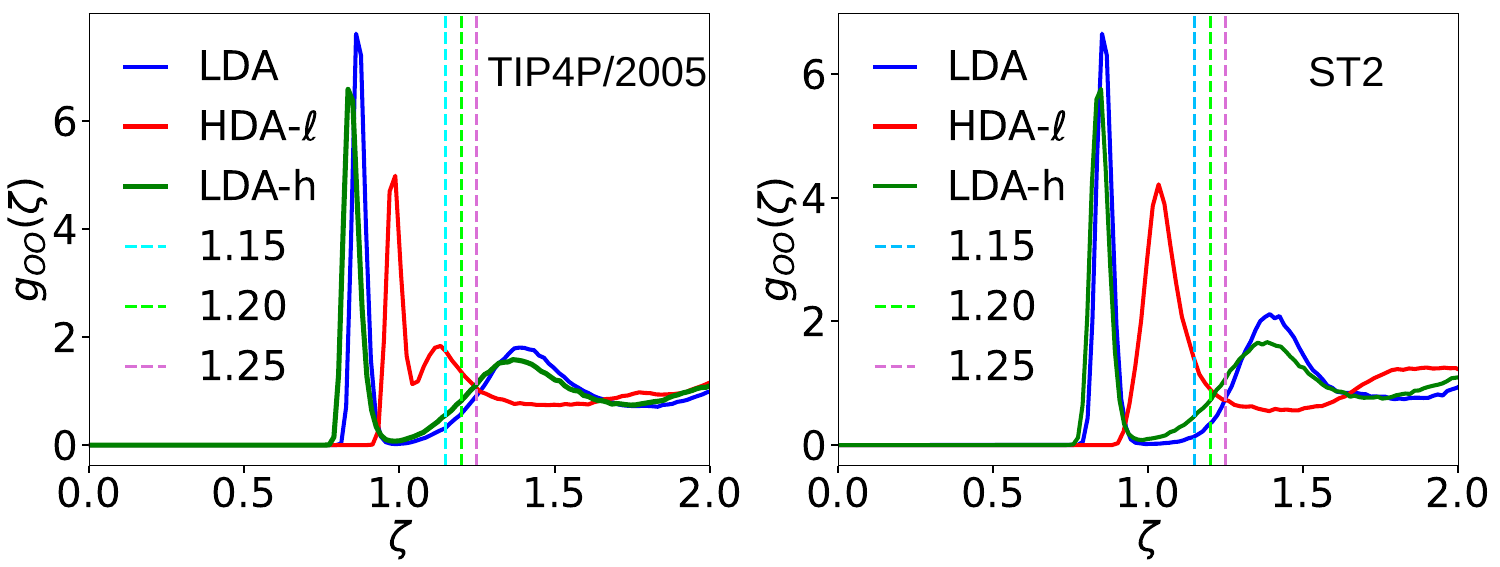}
    \caption{The scaled oxygen-oxygen radial distribution function, $g_{OO}(\zeta)$ ($\zeta \equiv r\rho^{1/3}$, where $\rho$ is water number density) for the LDA, HDA-$l$, and LDA-$h$ phases. The vertical lines represent different scaled radial cutoff distances used to define the neighbor list.}
    \label{rescaled_rdf}
\end{figure}

\begin{figure} [htbp]
    \centering
    \includegraphics[scale=0.45]{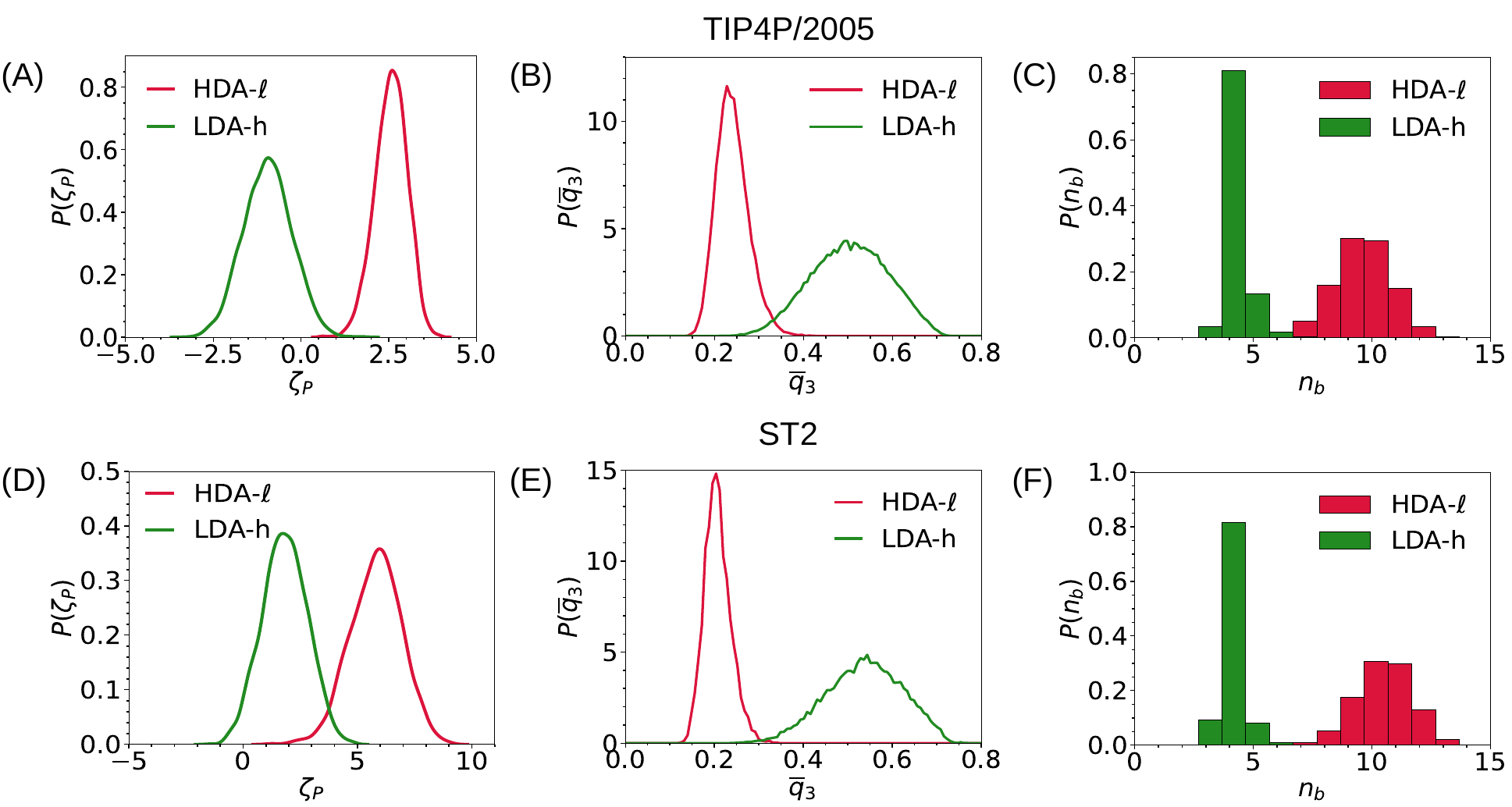}
    \caption{The local order parameter ($\zeta_{\rm p}$, $\overline{q}_3$, and $n_{\rm b}$) distribution for the HDA-$l$ and LDA-$h$ phases for both the water models --- TIP4P/2005 and ST2 is shown. The scaled radial cutoff distance of $1.2$ is used to define the neighbor list for the computation of $\overline{q}_3$.}
    \label{ql_dist}
\end{figure}

\begin{figure}
    \centering
    \includegraphics[scale=0.55]{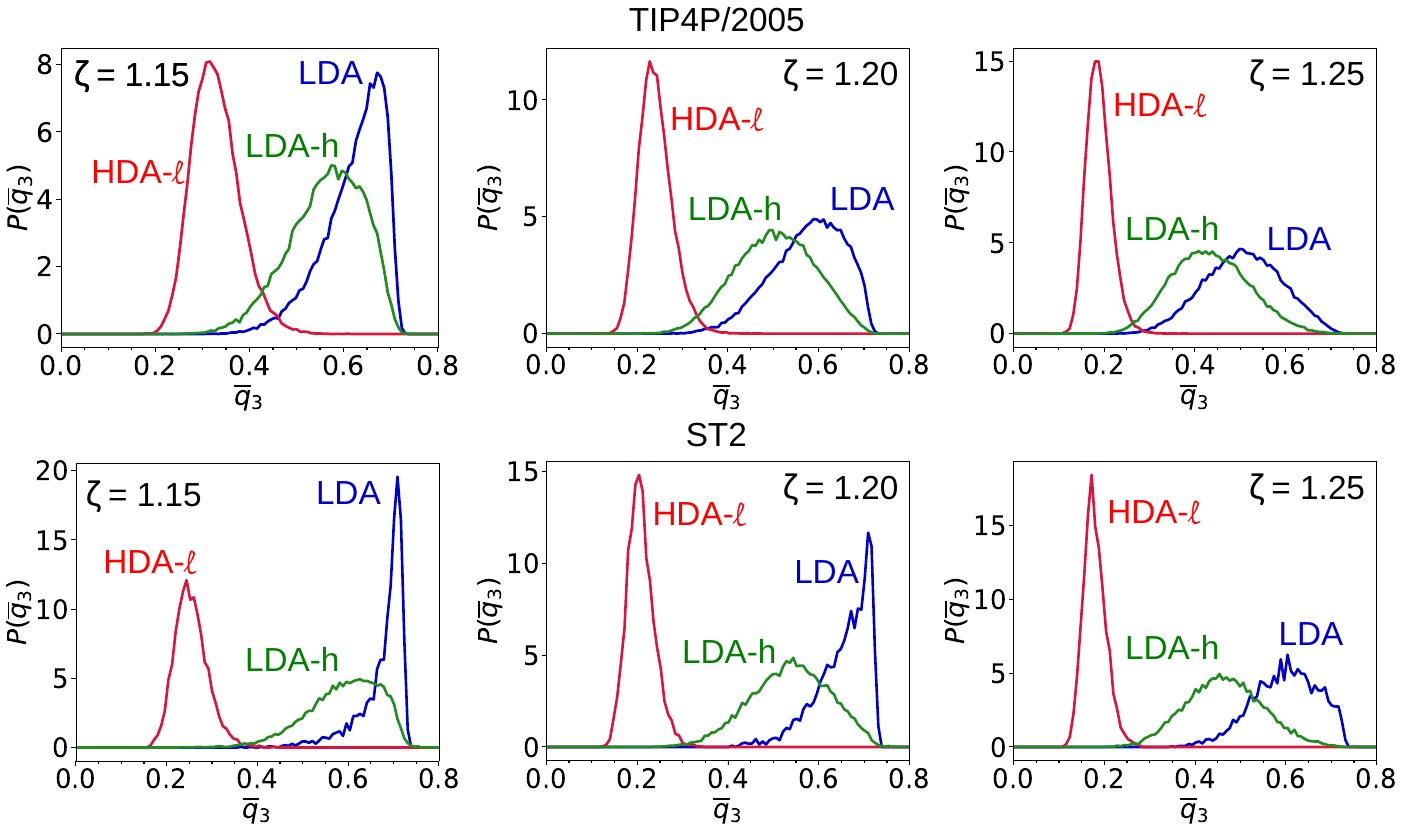}
    \caption{The dependence of the local $\overline{q}_3$ distribution on the choice of the scaled radial cutoff distance ($\zeta_{\rm c}$) used to define the neighbor list for the LDA, HDA-$l$, and LDA-$h$ phases.}
    \label{q3_cutoff}
\end{figure}

\begin{figure}
    \centering
    \includegraphics[scale=0.5]{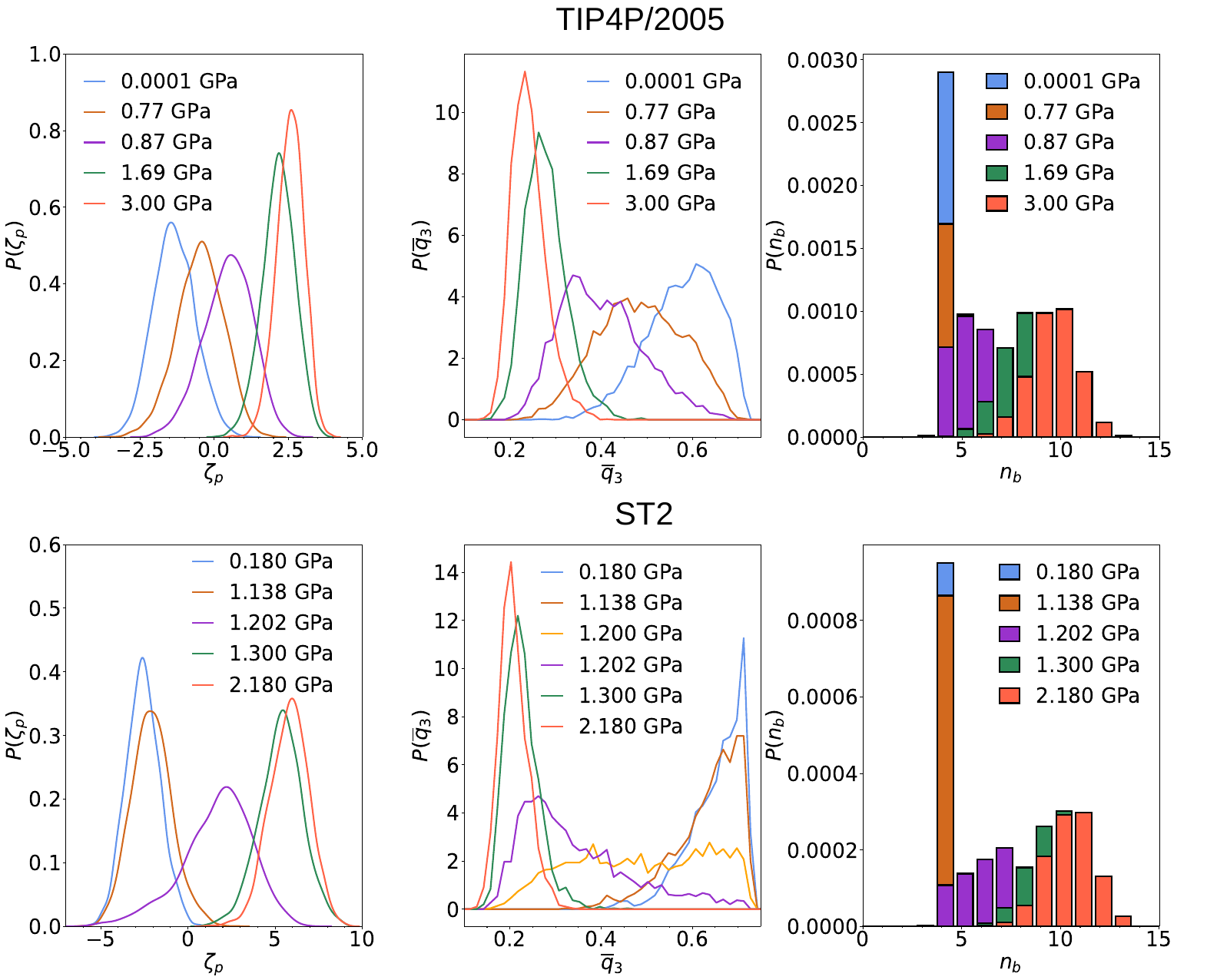}
    \caption{The distribution of the $\zeta_{\rm p}$, $\overline{q}_3$, and $n_{\rm b}$ order parameters at different pressures during the LDA to HDA-$l$ transition for the TIP4P/2005 (top) and ST2 (bottom) water. For the ST2 water, the order parameter distributions show a drastic change on changing $P$ near $P^*$ compared to the TIP4P/2005 water. This is not surprising as the transition in the density-pressure plane for the ST2 is much sharper compared to the TIP4P/2005 water (see Fig.~\ref{rho_e}A) in the main text.}
    \label{opdist}
\end{figure}

\begin{figure*}[t!]
    \centering
    \includegraphics[scale=0.71]{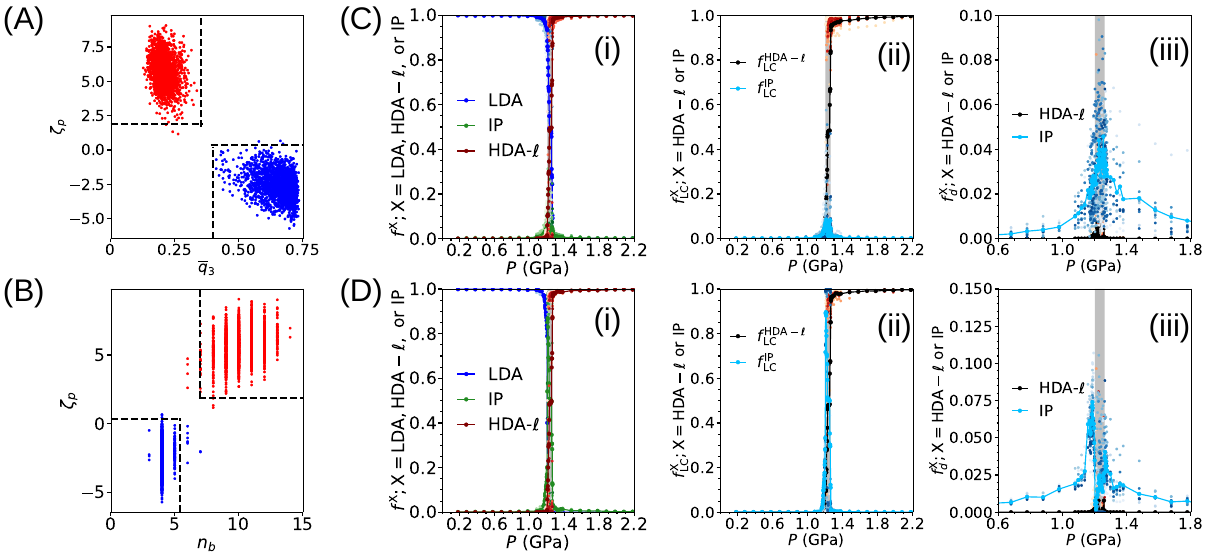}
    \caption{The LDA to HDA-$l$ phase transition pathway in the cluster-size space for the ST2 water. The LDA, HDA-$l$, and pre-ordered intermediate phase (IP)-like local environments are identified using a combination of two order parameters: $\zeta_{\rm p}$ and $\overline{q}_3$ (A); and $\zeta_{\rm p}$ and $n_{\rm b}$ (B).  The dashed lines represent the order parameter boundaries that best separate the LDA and HDA-$l$ along respective directions. The clustering analysis results on structural inhomogeneities classified using (A) and (B) are shown in (C) and (D), respectively. The $P$-dependent fraction of (i) the LDA-like, IP, and HDA-$l$-like particles, (ii) the HDA-$l$-like and IP particles that are part of the respective largest cluster ($f_{\rm LC}^{\rm X}$, with X = HDA-$l$ or IP), and (iii) the HDA-$l$ and IP particles that are not part of the respective largest cluster ($f_d^{\rm X}$, with X = HDA-$l$ or IP) is reported. The scatter points represent results from all $10$ independent trajectories and the solid lines indicate the average behavior. The shaded region is bounded by the smallest and largest $P^*$ values evaluated for each independent compression trajectory of the LDA phase.} 
   \label{clust_st2}
\end{figure*}

\begin{figure*}[t!]
    \centering
    \includegraphics[scale=0.6]{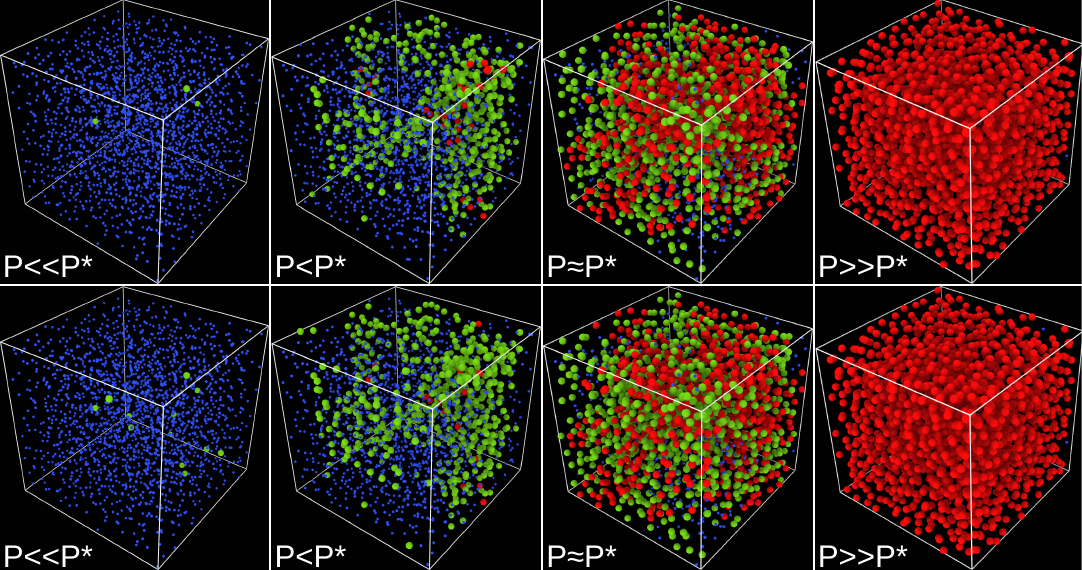}
    \caption{Representative snapshots of the ST2 water system at different pressures along isothermal compression-induced LDA to HDA-$l$ transition. The LDA-like, IP, and HDA-like particles are represented by blue, green, and red spheres, respectively. We note enhanced structural heterogeneities in the vicinity of $P^*$.} 
   \label{snaps_st2}
\end{figure*}

\begin{figure}[t!]
    \centering
    \includegraphics[scale=0.52]{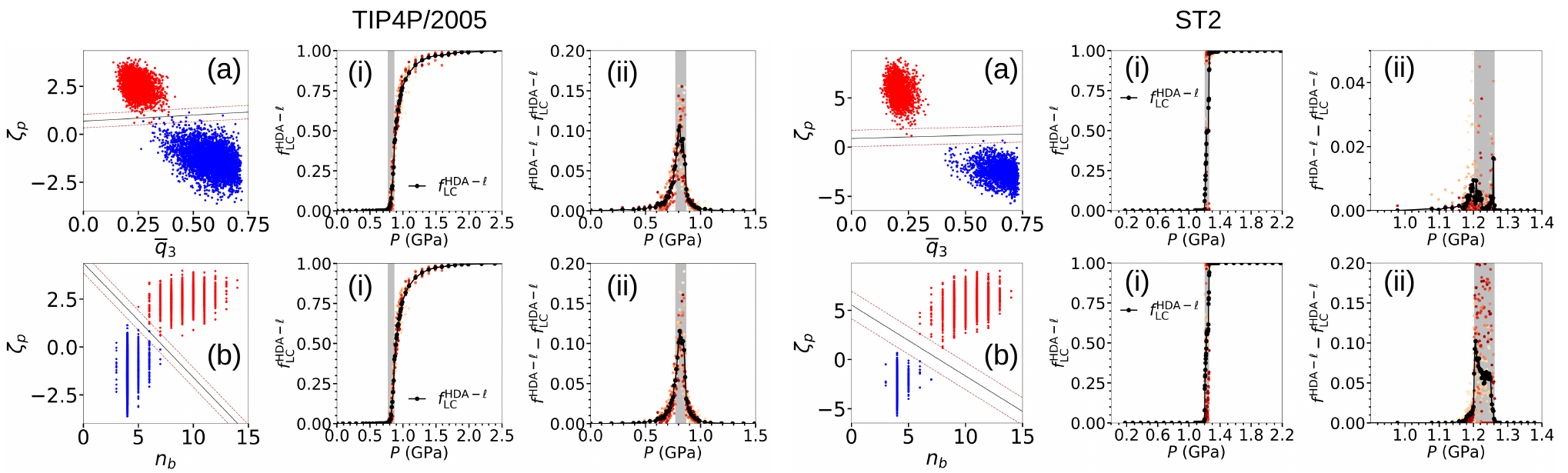}
    \caption{The LDA to HDA-$l$ phase transition pathway in the cluster-size space for the TIP4P/2005 and ST2 water using binary phase classification based on two order parameters: $\zeta_{\rm p}$ and $\overline{q}_3$ (a); and $\zeta_{\rm p}$ and $n_{\rm b}$ (b). (i) The fraction of HDA-$l$-like particles that are part of the largest HDA-$l$ cluster.  The scatter points represent results from all $10$ independent trajectories. (ii) The fraction of HDA-$l$ particles that are not part of the largest HDA-$l$ cluster $(f_d^{{\rm HDA}\mbox{-}l})$. The shaded region is bounded by the smallest and largest $P^*$ values evaluated for each independent isothermal compression trajectory of the LDA phase for both the water models. The $f_d^{{\rm HDA}\mbox{-}l}$ shows a maximum near $P^*$ for both the models, suggesting a collective spinodal-like transition mechanism, especially for the TIP4P/2005 water.} 
   \label{rhol_clust_binary}
\end{figure}

\begin{figure}[hb!]
    \centering
    \includegraphics[scale=0.6]{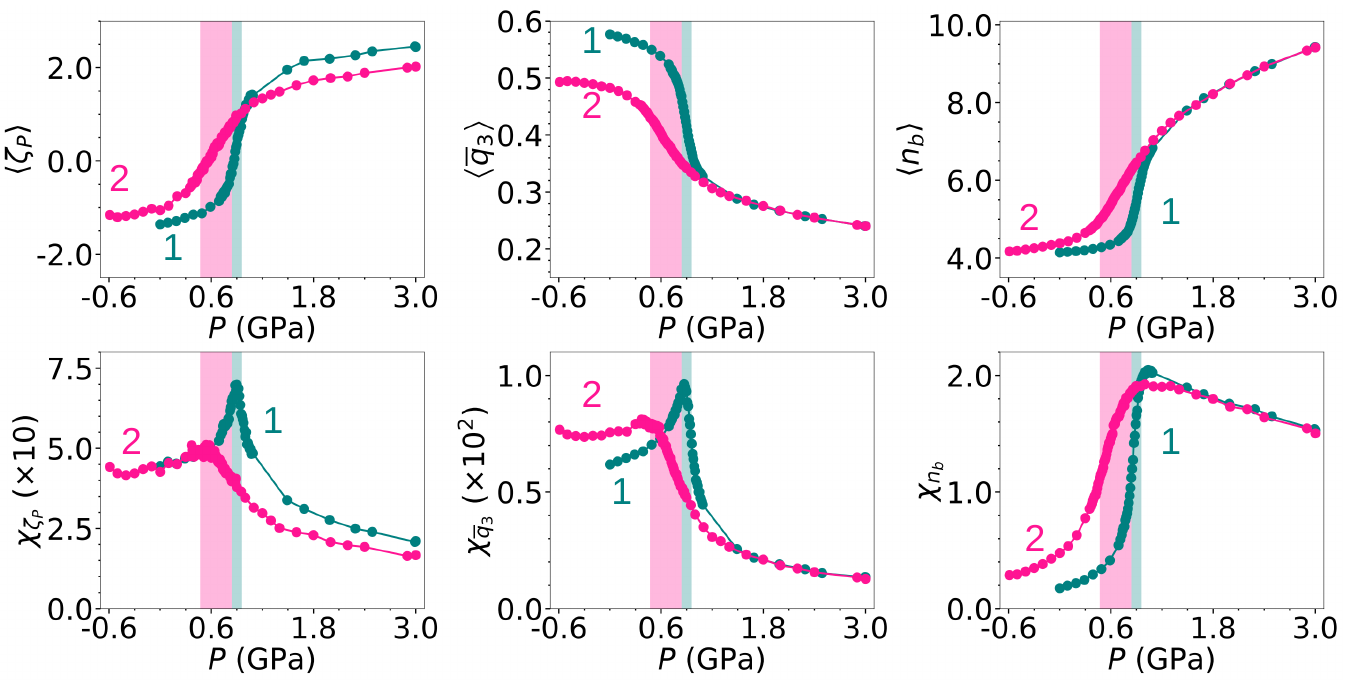}
    \caption{The average and the corresponding susceptibility of the local order parameters --- $\zeta_{\rm p}$, $\overline{q}_3$, and $ n_{\rm b}$ for the TIP4P/2005 water. The shaded regions are bounded by the lowest and highest $P^*$ values obtained from the ten independent simulations for each compression and decompression-induced phase transition. The teal and pink colors represent Path $1$ and Path $2$ (indicated by $1$ and $2$), respectively.}
    \label{avg_dist_sus_memory}
\end{figure}

\begin{figure}
    \centering
    \includegraphics[scale=0.6]{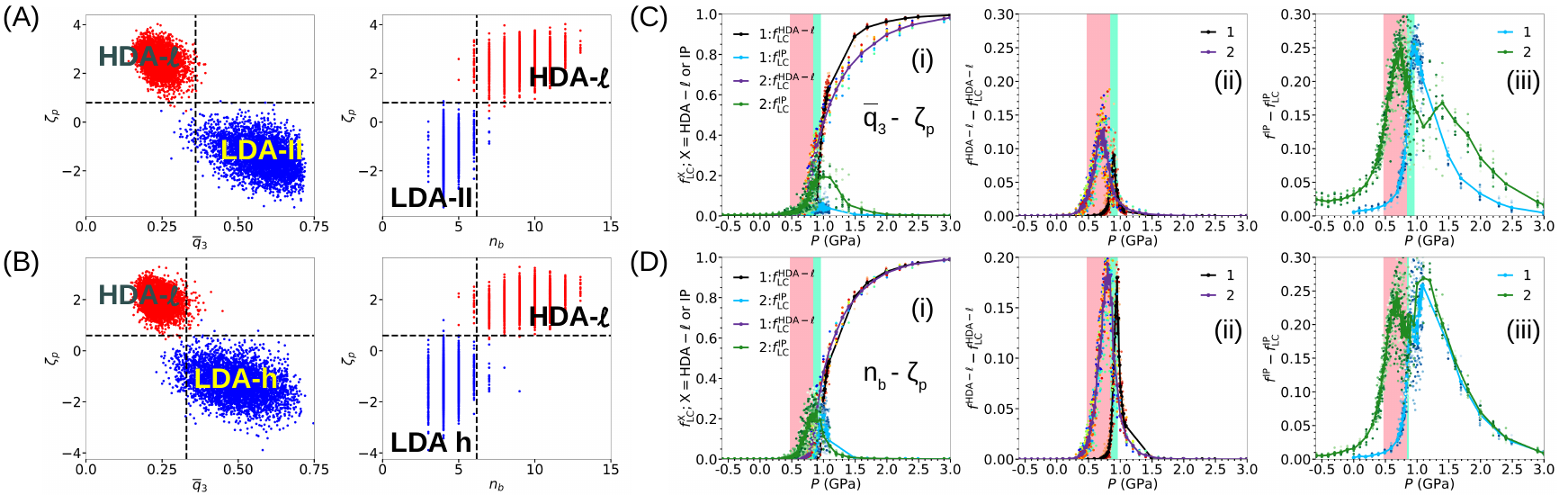}
    \caption{The LDA-like, IP, and HDA-$l$-like local environments are identified using a combination of two order parameters: $\zeta_{\rm p}$ and $\overline{q}_3$; and $\zeta_{\rm p}$ and $n_{\rm b}$ for paths $1$ (A) and $2$ (B). The phase transition pathway in the cluster size space using the phase classification for path $1$ (C) and path $2$ (D). (i) The growth of the HDA-$l$-like particles which are part of the largest HDA-$l$ cluster, along with the growth of the IP particles that are part of the largest IP cluster is shown. (ii) The fraction of the HDA-$l$-like particles that are not part of the largest HDA-$l$ cluster is shown. (iii) The fraction of the IP particles which are not part of the largest IP cluster is shown. The shaded green region is the region bounded by the lowest and highest $P^*$ values for path $1$, and the shaded pink region is the same for path $2$. We note the transition mechanism is qualitatively the same for both the paths.}
 \label{cluster_memory}
\end{figure}

\end{document}